\documentstyle{elsart}
\begin{document}
\begin{frontmatter}
\title{Monopoles and Gluon Fields in QCD
 \\ in the Maximally Abelian Gauge}
\author{Hiroko Ichie\thanksref{Email2}}
\thanks[Email2]{E-mail: ichie@th.phys.titech.ac.jp}
\address{
Department of Physics, Tokyo Institute of Technology \\
Ohokayama 2-12-1, Meguro, Tokyo 152-8551, Japan 
} 
\author{Hideo Suganuma\thanksref{Email1}}
\thanks[Email1]{E-mail: suganuma@rcnp.osaka-u.ac.jp}
\address{
Research Center for Nuclear Physics (RCNP), Osaka University \\
Mihogaoka 10-1, Ibaraki, Osaka 567-0047, Japan }
\begin{abstract} 
We study monopoles and gluon fields in QCD in the maximally 
abelian (MA) gauge in the context of the dual superconductor 
picture for confinement. 
In the abelian gauge, unit-charge magnetic monopoles appear,
but multi-charge monopoles do not in general cases. 
The appearance of the monopole is studied using the gauge-connection 
formalism in relation to the SU($N_c$) singular gauge transformation. 
The relevant role of off-diagonal gluons is found 
for the appearance of monopoles in the abelian gauge in QCD. 
We study the gluon-field properties around the monopole 
in the MA gauge in terms of the action density using the lattice QCD. 
The monopole provides infinitely large field fluctuations 
in the abelian sector. 
In the MA gauge, off-diagonal gluons are strongly suppressed but 
largely remain around the monopole, which 
indicates the effective size and the structure of monopoles. 
We find the large cancellation between the abelian part 
and the off-diagonal part of the action density 
around the monopole in the MA gauge. 
Owing to this cancellation, the monopole can appear in QCD 
without large cost of the QCD action. 
Finally, we generalize the framework of the abelian projection, 
{\it i.e.} the extraction of the abelian gauge manifold from QCD, 
by introducing the `gluonic Higgs field' 
$\vec \phi_D[A_\mu(x)]$ defined from 
the ${\rm SU}(N_c)$ covariant derivative $\hat D_\mu$. 
By way of $\vec \phi_D[A_\mu(x)]$, 
the maximally abelian projection can be performed
in the gauge-covariant manner 
without the notion of gauge fixing in principle. 
\end{abstract}
\end{frontmatter}

\section{Monopoles and Confinement in QCD}

In the classical and quantum field theories \cite{itzykson,cheng}, 
there occasionally 
appears the topological object as the interesting collective 
degrees of freedom, reflecting 
the nontrivial topology of the 
fiber-bandle. For instance, the Abrikosov vortex \cite{abrikosov}
is experimentally 
observed in the type-II superconductor, and 
the instanton \cite{diakonov,shuryak} in the Euclidean Yang-Mills theory is 
observed in the lattice QCD simulation using the cooling 
method\cite{hashimoto,diacomo1}. 
The  magnetic monopole \cite{thooftM,polyakovM} is also the 
interesting topological object 
predicted in the grand unified theory (GUT). 
%
%
Here, the magnetic monopole was firstly introduced 
by Dirac more than 50 years ago from the consideration of 
the duality of the Maxwell equation, and the Dirac monopole \cite{dirac}
can naturally explain the electric-charge quantization. 
However, the Dirac monopole cannot be an extended object 
so as to make the Dirac string invisible, and 
such a point-like monopole is not allowed in QED, 
because it provides the divergence of the QED action. 
In 1974, however,  the magnetic monopole was well formulated as the 
't~Hooft-Polyakov monopole\cite{thooftM,polyakovM}, 
which is the topological object 
in the nonabelian Higgs theory with the 
compact and at most semi-simple group.

Also in the $N=2$ supersymmetric (SUSY) QCD, 
the soliton-like monopole is recognized as an essential degrees of 
freedom in the strong-coupling region, 
and its condensation in the infrared region 
provides the dual-superconductor picture for confinement in 
SUSY-QCD\cite{seiberg}.
As for QCD, however, it seems difficult to introduce the 
(color-)magnetic monopole because of the absence of the Higgs field. 
Nevertheless, the introduction of the monopole degrees of freedom 
is desired for the physical interpretation of 
the confinement phenomena in QCD.


In 1970's, Nambu, 't~Hooft  and Mandelstam 
proposed an interesting idea of the electric confinement 
by magnetic-monopole condensation, and tried the physical 
interpretation of 
quark confinement using the dual version of the 
superconductivity\cite{nambu,thoa,mandelstam}.
In the ordinary superconductor, 
Cooper-pair condensation leads to the Meissner effect, 
and the magnetic flux is excluded or
squeezed like a 
quasi-one-dimensional tube as the Abrikosov vortex.
On the other hand, 
from the Regge trajectory of hadrons and the lattice QCD\cite{rothe}, 
the confinement force between the color-electric charge is 
characterized by 
the universal physical quantity of 
the string tension $\sigma \simeq 1{\rm GeV/fm}$, 
and is brought by one-dimensional squeezing of 
the color-electric flux
\cite{haymaker}
in the QCD vacuum. 
Hence, from the above similarity on the one-dimensional flux 
squeezing, the QCD vacuum was regarded as the dual version 
of the superconductor.
In this dual-superconductor picture for the QCD vacuum, 
the squeezing of the color-electric flux between quarks 
is realized by the dual Meissner effect
as the result of condensation of color-magnetic monopoles.
However, there are two 
large gaps between QCD and the dual 
superconductor picture. 
\begin{enumerate}
\item
This picture is based on the abelian gauge theory 
subject to the 
Maxwell-type equations, where electro-magnetic duality is manifest, 
while QCD is a nonabelian gauge theory, described with the electric 
variables (quarks and gluons).
\item
The dual-superconductor scenario requires condensation of 
magnetic \\  monopoles as key concept, while 
QCD does not have such a monopole as 
the elementary degrees of freedom. 
\end{enumerate}
\indent As the connection between QCD and the dual superconductor scenario, 
't~Hooft proposed concept of the abelian gauge fixing\cite{thooft}, 
the partial gauge
 fixing which is defined by  diagonalizing a suitable 
gauge-dependent variable as $\phi[A_\mu(x)]$.
The abelian gauge fixing reduces QCD into an abelian gauge
theory, where the off-diagonal element of the gluon field behaves as a 
charged matter field. 
As a remarkable fact in the abelian gauge, color-magnetic monopoles appear
as topological objects corresponding to 
the nontrivial homotopy group $\Pi_2( {\rm SU}(N_c)/ {\rm U(1)}^{N_c-1}) =
{\bf Z}^{ N_c-1}_\infty$.
Here, assuming  abelian dominance \cite{ezawa}, 
which means
that the only abelian gauge fields with monopoles 
would be essential for the description of the
nonperturbative QCD,
the off-diagonal gluon elements are dropped off,  
which is called the abelian projection. 
Thus, by the abelian gauge fixing and the abelian projection, 
QCD is reduced into  abelian projected QCD (AP-QCD), which is an 
abelian gauge theory including monopoles.
If the monopole condenses, 
the scenario of color confinement by the dual Meissner effect  would be 
a realistic picture for confinement in QCD 
\cite{maedan,suganuma,sasaki,umisedo,ichie6,ichie1,ichie2,atanaka,kondo}.

Recent lattice QCD simulations show strong evidence on this dual 
Higgs theory for the nonperturbative QCD in the maximally abelian 
(MA) gauge\cite{diacomo,poly}.
The MA gauge is the abelian gauge where the off-diagonal gluon  
is minimized by the gauge transformation.
In the MA gauge, the physical information of the gauge configuration is 
concentrated into the diagonal components as well as possible.
The lattice QCD studies indicate {\it abelian dominance}
\cite{ezawa,ichiead,suganuma1} 
that the string tension\cite{yotsuyanagi,hioki,bali} 
and the chiral condensate \cite{miyamura,woloshyn} are almost described only 
by abelian variables in the MA gauge \cite{kronfeld,schierholz}.
In the lattice QCD, 
{\it monopole dominance} is also observed such that 
only the monopole part in the abelian variable 
contributes to the nonperturbative QCD in the MA gauge \cite{bali,miyamura}.
Thus, the lattice QCD studies also suggest the dominant role of 
abelian variables and monopoles 
for the nonperturbative QCD in the MA gauge.

In the MA gauge, 
 AP-QCD neglecting 
the off-diagonal gluon component almost reproduces 
the essence of the nonperturbative QCD, 
although AP-QCD is an abelian gauge theory like QED.
One may speculate that the strong-coupling nature 
leads to the similarity between AP-QCD and QCD, 
because the gauge coupling $e$ in AP-QCD \cite{kondo} is 
the same as that in QCD in the lattice simulation. 
However, the strong-coupling nature would not be enough 
to explain the nonperturbative feature, 
because, if monopoles are eliminated from AP-QCD, 
nonperturbative features are lost in 
the remaining system called as the photon part, 
although the gauge coupling $e$ is the same as that in QCD.

For further understanding, 
we compare the theoretical structure of QCD, 
AP-QCD and QED in terms of the gauge symmetry 
and the relevant degrees of freedom, as shown in Fig.\ref{APQCD}. 
As for the interaction, the linear confinement potential arises 
both in QCD and both in AP-QCD, 
while only the Coulomb potential appears in QED.
On the symmetry, QCD has nonabelian gauge symmetry,
while both AP-QCD and QED have abelian gauge symmetry. 
The obvious difference between QCD and QED is existence of  
off-diagonal gluons. 
On the other hand, the difference between AP-QCD and QED is 
 existence of the monopole, since the magnetic monopole does not 
exist in QED because of the Bianchi identity. 
This indicates the close relation between monopoles and off-diagonal 
gluons.
In particular, off-diagonal gluon components play a crucial role 
for existence of the monopole in QCD as shown below, 
and the monopole itself is expected to play an alternative role of 
off-diagonal gluons for the confinement.

Here, we consider what is the QCD-monopole 
in comparison with the Cooper pair in the superconductivity. 
In the field theoretical aspect, 
the essence of the superconductivity is understood 
as the ordinary Higgs mechanism by Cooper-pair condensation, 
although the underlying electron-phonon interaction plays 
relevant role for the creation of the Cooper pair. 
The composite Cooper-pair field plays the role of the Higgs field 
and is the essential degrees of freedom for the superconductivity. 
Similarly, 
the monopole field to be condensed in the nonperturbative QCD vacuum 
can be regarded as a kind of composite or collective degrees of freedom 
relevant for the nonperturbative phenomena in QCD, 
since QCD includes quarks and gluons as the elementary field only. 
Different from the simple compositicy of the Cooper pair, 
the QCD-monopole appears as a topological object 
relating to the singularity of the gauge field in the abelian gauge, 
and would be described as a complicated field composed by gluons. 
Then, the feature of the structure of the monopole in QCD is to be 
clarified using the field-theoretical framework including the lattice 
QCD.

In this paper, we study the properties of the monopole appearing 
in QCD in the MA gauge in terms of the gluon field around it 
both in the analytical framework and in the lattice QCD. 
In section 2, 
we study the general argument of the abelian gauge fixing 
considering the global Weyl transformation. 
In the abelian gauge, the monopole appears from 
the hedgehog configuration of the gluonic Higgs field
through the SU($N_c$) singular gauge transformation. 
We clarify the appearance of monopoles in terms of the 
gauge connection with respect to the singularity 
of the SU($N_c$) gauge transformation. 
We extract the abelian gauge field and the 
monopole current in the lattice formalism. 
In section 3, we study the MA gauge fixing in detail 
in terms of the abelian projection rate, and 
propose the transparent definition of the MA gauge using the 
covariant derivative. 
The generalization of the MA gauge is also attempted. 
In section 4, 
we examine the gluon properties around the monopoles 
in the MA gauge using the lattice QCD simulation, 
with considering the role of the off-diagonal gluons. 
In section 5, 
we introduce the `gluonic Higgs field' extracted from 
the ${\rm SU}(N_c)$ gauge connection $\hat D_\mu$, and 
formulate the abelian projection in the gauge-covariant manner without 
explicit use of gauge fixing.
Section 6 is devoted to summary and concluding remarks.

\section{Appearance of Monopoles in the Abelian Gauge}

\subsection{Abelian Gauge Fixing and Residual Symmetries}

The dual superconductor picture for confinement phenomena 
is based on the abelian gauge theory including monopoles, 
and the 't~Hooft abelian gauge fixing\cite{thooft} 
is  key concept for the connection from QCD 
to such an abelian gauge theory.
In this section, we study the abelian gauge fixing 
considering the role of the global Weyl symmetry 
and show the appearance of monopoles in the singular SU($N_{c}$) 
gauge transformation.
 
The abelian gauge fixing is
the partial gauge fixing which remains the abelian gauge symmetry, and
is realized by the diagonalization of an 
SU($N_c$) gauge-dependent variable 
as $\phi[A_\mu(x)] = \phi^a T^a \in su(N_c)$
using the SU($N_c$) gauge transformation.
In the abelian gauge, $\phi[A_\mu(x)]$ plays the 
similar role of the Higgs field \cite{ichiec} in the determination of 
the gauge fixing, and then we call $\phi[A_\mu(x)]$
as the `gluonic Higgs field'. 

Without loss of generality, we consider the case of 
the hermite variable $\phi[A_\mu(x)]$ which 
obeys the adjoint gauge transformation.
Then,
$\phi(x)$ is transformed as
\begin{eqnarray}
\phi(x) = \phi^a(x) T^a 
\rightarrow \phi^\Omega(x) & =  & \Omega(x) \phi(x) \Omega^{\dagger}(x) \\
& \equiv  & {\rm diag}(\lambda^1(x), \cdots, \lambda^{N_c}(x))
\equiv \vec H \cdot \vec \lambda(x)  \nonumber 
\end{eqnarray}
using a suitable SU$(N_{c})$ gauge function 
$\Omega(x) = {\rm exp} \{ i \xi^a (x) T^a \} \in$ SU($N_{c}$).
Here, each diagonal component $\lambda^{i}(x)$ 
($i$=1,$\cdots$, $N_c$) is to be real for the 
hermite variable $\phi[A_\mu(x)]$.
The space-time point $x$ satisfying $\lambda^i(x) = \lambda^j(x)$
is called as the `degeneracy point' and reflects the singular structure 
of $\phi(x)$.
Particularly for the SU(2) case, one finds
\begin{eqnarray}
\phi(x) \equiv \phi^a (x) \frac{\tau^a}{2} \rightarrow 
\phi^\Omega (x) = \Omega \phi \Omega ^\dagger = \lambda(x) 
\frac{\tau^3}{2}
\end{eqnarray}
with $\lambda(x) = \pm\{\phi^1(x)^2 + \phi^2(x)^2+\phi^3(x)^2\}^{1/2}$.
In the abelian gauge, the SU($N_{c}$) gauge symmetry is reduced into 
the U(1)$^{N_{c}-1}$ gauge symmetry, which  
corresponds to the gauge-fixing ambiguity. 
The variable $\phi(x)$ is diagonalized to 
$\vec H \cdot \vec \lambda(x)$ also 
by the gauge function $\Omega^\omega (x) \equiv \omega(x)\Omega(x)$ with 
$\omega(x) = e^{- i \, \vec H \cdot  \vec \varphi} \in$
U(1)$^{N_{c}-1}$, 
\begin{eqnarray}
\phi(x) \rightarrow  \Omega^\omega(x) \phi(x)
\Omega^{\omega\dagger}(x) = \omega(x) \vec  H \cdot \vec \lambda(x)
\omega^{\dagger}(x) = \vec H \cdot  \vec \lambda(x),
\end{eqnarray}
and therefore U(1)$^{N_{c}-1}$ abelian gauge symmetry remains 
in the abelian gauge.
Hence, the diagonal and off-diagonal gluon components play 
the different role in the abelian gauge;
the diagonal gluon remains to be the abelian gauge field,
while the off-diagonal gluon behaves as the charged matter 
like $W_{\mu}^{\pm}$ in the Standard Model.
In the continuum theory, the abelian projection, the extraction of the 
abelian gauge manifold, is defined by the simple replacement as
\begin{eqnarray}
A_{\mu} \equiv A^a_{\mu} T^a \,\, \in \,\, su(N_{c}) \,\,\, \rightarrow
{\cal A_{\mu}} \equiv {\rm tr} (A_{\mu} \vec H) \cdot \vec H \in 
u(1)^{N_{c}-1} 
\end{eqnarray}
after the abelian gauge fixing.

In the abelian gauge, there also remains the global Weyl symmetry as a 
`relic' of the nonabelian theory \cite{ichiead,suganuma4,suganuma2}.
The Weyl symmetry is the permutation group ${\bf P}_{N_{c}}$
corresponding to the 
permutation of the $N_{c}$ bases in the fundamental representation.
For the SU(2) case, the Weyl transformation is expressed as
the constant off-diagonal SU(2) matrix
\begin{eqnarray}
W  & \equiv  & e^{ i \{\frac{\tau_{1}}{2} \cos \alpha +
\frac{\tau_{2}}{2} \sin \alpha \} \pi} = i(\tau_{1} \cos \alpha + 
\tau_{2} \sin \alpha)  \\ \nonumber & = & 
i \left( {\matrix{
0  &
e^{-i\alpha} \cr
e^{i\alpha} &
0
}} \right) 
\in {\bf P}_{N_{c}=2} \,\, \subset \,\, \mbox{\rm SU}(2)
\label{eq:weyl}
\end{eqnarray}
 with 
$\alpha \in {\bf R}$ fixed. 
In the abelian gauge, the variable $\phi(x)$ is also diagonalized by using
$\Omega^W(x) \equiv W \Omega(x)$,
\begin{eqnarray}
\phi(x) \rightarrow  \Omega^{W}(x) \phi(x)
\Omega^{W{\dagger}}(x) = W \lambda(x) \frac{\tau^3}{2} W ^{\dagger}
= - \lambda(x) \frac{\tau^3}{2}.
\end{eqnarray}
Here, the sign of $\lambda(x)$, or the order of 
the diagonal component $\lambda^{i}(x)$, is globally changed 
by the Weyl transformation. 
It is noted that 
the sign of the U(1)$_3$ gauge field 
${\cal A}_\mu \equiv A^3_\mu \frac{\tau_3}{2}$ is also globally changed
under the Weyl transformation,
\begin{eqnarray}
{\cal A}_\mu \rightarrow {\cal A}_\mu^{W} = W  A^3_\mu \frac{\tau_3}{2} W^\dagger = 
- A^3_\mu \frac{\tau_3}{2} = -{\cal A}_\mu.
\end{eqnarray}
Therefore, all the sign of the abelian field strength  ${\cal 
F}_{\mu\nu}$ as  $ (\partial \wedge {\cal A})_{\mu\nu}$,
electric and  magnetic charges are also globally changed:
\begin{eqnarray}
{\cal F}_{\mu\nu}  & \equiv & F_{\mu\nu}\frac{\tau_3}{2}  \rightarrow   
{\cal F}_{\mu\nu}^{W} = W  {\cal F}_{\mu\nu}  W^\dagger 
= - {\cal F}_{\mu\nu},  \nonumber \\
j_\mu & \equiv & \partial^\alpha {\cal F}_{\alpha\mu}   
\rightarrow   j_\mu^{W}=-j_\mu, \nonumber \\
k_\mu & \equiv & \partial^\alpha {}^*{\cal F}_{\alpha\mu}   
\rightarrow   k_\mu^{W}=-k_\mu.
\end{eqnarray}
In the abelian gauge, the absolute sign of the electric 
and the magnetic charges is settled, only when the Weyl symmetry 
is fixed by the additional condition.
When $\phi[A_{\mu}(x)]$ obeys the simple adjoint gauge transformation 
like the nonabelian Higgs field, the global Weyl symmetry can be 
easily fixed by imposing the additional gauge-fixing condition 
as $\lambda (x) \ge 0$ for SU(2), or 
the ordering condition of the diagonal components $\lambda^{i}(x) $ 
in $\vec H \cdot \vec \lambda$ 
as $\lambda^{1}(x) \ge..\ge \lambda^{N_c}(x)$ for the SU($N_{c}$) case. 
As for the appearance of monopoles in the abelian gauge, 
the global Weyl symmetry ${\bf P}_{N_c}$ is not relevant, because 
the nontriviality of the homotopy group is 
not affected by the global Weyl symmetry. 
However, the definition of the magnetic monopole charge, 
which is expressed by the nontrivial dual root 
of ${\rm SU}(N_c)_{\rm dual}$ \cite{ezawa}, 
is globally changed by the Weyl transformation.

\subsection{Monopoles and the Hedgehog Configuration of Gluonic
Higgs Field}

The abelian gauge fixing, which reduces QCD into an abelian gauge
theory, is realized by the diagonalization of a gluonic Higgs field
$\phi[A_\mu(x)]$. 
In the continuum theory of QCD, the continuous field $A_\mu(x)$ can be
taken to be regular everywhere in a suitable gauge as the Landau 
gauge, and then
$\phi[A_\mu(x)]$ is expected to be a regular function almost everywhere.
In the abelian gauge, however, there appears the singular point,
where the gauge function to diagonalize $\phi[A_\mu(x)]$ is not 
uniquely determined even for the off-diagonal part, and  such a 
singular point 
leads to the
appearance of the monopole.
Here, let us consider the appearance of QCD-monopoles in the abelian gauge 
in terms of the singularity in the gauge transformation\cite{suganuma}.
For the gluonic Higgs field 
$\phi(x)$ obeying the adjoint transformation,
the monopole appears at the `degeneracy point' of 
the diagonal elements of $\vec H \cdot \vec \lambda(x) = 
diag(\lambda^{1}(x),\lambda^{2}(x),\cdots, \lambda^{N_{c}}(x))$
after the abelian gauge fixing: 
$(i,j)$-monopole appears at the point satisfying 
$\lambda^{i}(x)=\lambda^{j}(x)$.
For the $(i,j)$-monopole, the SU(2) subspace relating  to $i$ and $j$
is enough to consider, so that the essential feature of the monopole 
can be understood in the SU(2) case without loss of generality.
Then, we consider the SU(2) case for simplicity. For the SU(2) case,
the diagonalized element of $\phi(x)$ are given by 
$\lambda = \pm(\phi_{1}^2 +\phi_{2}^2 +\phi_{3}^2)^{1/2}$, and hence 
the `degeneracy point' satisfies the condition $\phi(x)= 0$, 
which is ${\rm SU}(2)$ gauge invariant. 
This gauge-invariant condition $\phi(x)=0$ can be regarded as 
the singularity condition on $\hat \phi(x)\equiv \phi(x)/|\phi(x)|$ 
with $|\phi(x)|\equiv (\phi^a(x)\phi^a(x))^{1/2}$. 
In fact, the `degeneracy point' in the abelian gauge 
appears as the singular point of $\hat \phi(x)$ like the center 
of the hedgehog configuration as shown in Fig.\ref{Higgs}(b) 
before the abelian gauge fixing.

Since the singular point on $\hat \phi(x)$ is to satisfy 
three conditions $\phi^1(x)=\phi^2(x)=\phi^3(x)=0$ simultaneously, the set of 
the singular point forms the point-like manifold in ${\bf R}^3$
or the line-like manifold in ${\bf R}^4$.
We investigate the topological nature near the singular point 
$({\bf x}_0,t)$ of $\hat \phi(x)$ for fixed $t$, i.e., 
$\phi({\bf x}_0,t)=0$\cite{suganuma}.
Using the Taylor expansion,
one finds 
\begin{eqnarray}
\phi({\bf x},t) = \phi^a({\bf x},t) \frac{\tau^a}{2} 
\simeq \tau^a C^{ab} ({\bf x}-{\bf x}_0)^b,
\end{eqnarray}
with $C^{ab} \equiv \frac12 \partial^{b} \phi^a({\bf x}_0,t) $.
In the general case, one can expect 
det$C \ne 0$, i.e., det$C > 0$ or
det$C < 0$, and the fiber-bandle $\phi^a({\bf x})$ can be deformed 
into the (anti-)hedgehog configuration
$\phi(\tilde {\bf x}) \simeq \pm \tau^a \tilde {\bf x}^a$
around the singular point ${\bf x}_0$
by using the continuous modification on the spatial 
coordinate ${\bf x}^a \rightarrow \tilde {\bf x}^a \equiv 
{\rm sgn}({\rm det}C) \cdot C^{ab} ({\bf x}-{\bf x}_0)^b$.
The linear transformation matrix $C$ can be written 
by a combination of the rotation $R$ and the dilatation 
of each axis $\lambda = diag(\lambda^1,\lambda^2,\lambda^3)$
with $\lambda^i > 0$ as $C = {\rm sgn}( {\rm det}C) R \lambda$. 
Here, topological nature is never changed by such a continuous deformation.
For det$C > 0$, the configuration $\phi({\bf x})$
can be continuously deformed
into the hedgehog configuration around  ${\bf x}_0$, $\phi( \tilde {\bf x})
\simeq \tau^a \tilde {\bf x}^a$, while,
for det$C < 0$,  $\phi({\bf x})$ can be continuously deformed
into the anti-hedgehog configuration, $\phi( \tilde {\bf x})
\simeq - \tau^a \tilde {\bf x}^a$.
Since det$C=0$ is the exceptionally special case and  det$C < 0$ is 
similar to det$C > 0$, we have only to consider the hedgehog 
configuration.
This hedgehog configuration around the singular point 
of $\hat \phi(x)$ corresponds to the simplest nontrivial topology of 
the nontrivial homotopy group 
$\Pi_2({\rm SU(2)/U(1)_3})=Z_\infty$, and 
the abelian gauge field has the singularity as the monopole 
appearing from the hedgehog configuration.

Using the polar coordinate $(r,\theta,\varphi)$ of $\tilde {\bf x}$,
the hedgehog configuration is expressed as
\begin{eqnarray}
\phi & =  & \tau^a \tilde {\bf x}^a= 
   r \sin \theta \cos \varphi \cdot \tau_1 
+  r \sin \theta \sin \varphi \cdot \tau_2
+  r \cos \theta              \cdot \tau_3 
\nonumber \\  
 & = &
r  
\left( {\matrix{
{\rm cos}{\theta}   &
e^{-i\varphi}{\rm sin} \theta \cr
e^{i\varphi} {\rm sin} \theta &
 -{\rm cos}\theta }
} \right),
\end{eqnarray}
and $\phi$ can be diagonalized by the gauge transformation with
\begin{eqnarray}
 \Omega^H = 
\left( {\matrix{
e^{i\varphi}{\rm cos}{\frac{\theta}{2}}  & {\rm sin}{\frac{\theta}{2}} \cr
-{\rm sin}{\frac{\theta}{2}} & e^{-i\varphi}{\rm cos}{\frac{\theta}{2}}}
} \right),
\label{eq:gauge-function}
\end{eqnarray}
where $\theta$, $\varphi$ denote the polar and the azimuthal angles, 
respectively.
Here, on the $z$-axis ($\theta=0$ or $\theta=\pi$), $\varphi$ is the 
`fake parameter', and the unique description does not allow the 
$\varphi$-dependence on the $z$-axis. However, at the positive region 
of $z$-axis, $\theta = 0$, $\Omega^H$ depends on $\varphi$ and is 
multi-valued as
\begin{eqnarray}
 \Omega^H = 
\left( 
{
\matrix{
e^{i\varphi} & 0 \cr
0 & e^{-i\varphi}
}
} 
\right). 
\end{eqnarray}
Such a multi-valuedness of $\Omega^H$ leads to the divergence in the 
derivative $\partial_\mu  \Omega^H$ at $\theta = 0$. 
In fact, $\partial_\mu  \Omega^H$  includes the singular part as
 $\cos \frac{\theta}{2}$ 
$(\nabla \varphi)_\varphi = \frac{\cos \frac{\theta}{2}}{r \sin \theta} 
\frac{\partial}{\partial \varphi} \varphi = \frac{1}{r \sin 
\frac{\theta}{2}}$, which  diverges
at $\theta=0$.
By the gauge transformation with $ \Omega^H$,
the variable $\phi$ becomes
$\phi^{\Omega} = \Omega  \phi  \Omega^\dagger = r \tau^3$, and
the gauge field is transformed as
\begin{eqnarray}
A_\mu  \rightarrow  A_\mu^{\Omega} = 
 \Omega  (A_\mu  - \frac{i}{e} \partial_\mu)  \Omega^\dagger.
\end{eqnarray}
For regular $A_\mu$, the first term $ \Omega A_\mu  \Omega^\dagger$
is regular, while 
$A^{\rm sing}_\mu \equiv 
- \frac{i}{e}  \Omega  \partial_\mu  \Omega^\dagger$
is singular and
the monopole appears in the abelian sector originating from the singularity
of $A^{\rm sing}_\mu $
\cite{suganuma}.
To examine the appearance of the monopole at the origin $\tilde {\bf 
x}=0$, we consider the magnetic flux
$\Phi^{\rm flux}( \theta )$
which penetrates the area inside the closed contour 
$c(r,\theta) \equiv \{ (r,\theta,\varphi)  | 0 \le \varphi< 2 \pi \} $.
One finds that
\begin{eqnarray}
\Phi^{\rm flux}( \theta ) 
& = & \int_c d{\bf x} \cdot {\bf A}^{\rm sing}  
= -\frac{i}{e} \int_c  d{\bf x} \Omega {\nabla}  \Omega^\dagger  
 \nonumber \\ 
& = & -\frac{i}{e} \int^{2\pi}_{0} d \varphi  \Omega 
\frac{\partial}{\partial \varphi}   \Omega^{\dagger} 
=-\frac{4\pi}{e} \cdot \frac{1 + \cos \theta}{2}  \frac{\tau_3}{2},
\end{eqnarray}
which denotes the 
magnetic flux of the monopole with the unit-magnetic charge 
$g = \frac{4 \pi}{e}$ with the Dirac string \cite{suganuma}.
Here, 
the direction of the Dirac string from 
the monopole can be arbitrary changed by the singular U$_{3}(1)$ gauge 
transformation, which can move $e^{i\varphi}$ in $\Omega^H$ from the 
$\tau_{3}$-sector to the off-diagonal sector.
In fact, 
the multi-valuedness of $\Omega$ is not necessary to be fixed 
in $\tau^3$-direction.
Nevertheless, 
the singularity in 
$\Omega \partial_{\mu} \Omega^\dagger$ appears only in the 
$\tau_{3}$-sector, and $\tau_{3}$-direction becomes special in the 
abelian gauge fixing.

The anti-hedgehog configuration of 
$\phi( \tilde {\bf x})
= - \tau^a \tilde {\bf x}^a$
provides a
monopole with the  opposite magnetic charge, because anti-hedgehog 
configuration is transformed to the hedgehog configuration by the 
Weyl transformation. 
Thus, the only unit-charge magnetic monopole appears
in the general case of det$ C \ne 0$. 
In principle, the multi-charge 
monopole can 
also appear when  det$C =0$, however, the condition is scarcely satisfied
in general,
because this exceptional case is realized only when four conditions
$\phi^1 = \phi^2 = \phi^3 = {\rm det}C = 0$ are simultaneously 
satisfied.  
To summarize, in the abelian gauge,
the unit-charge magnetic monopoles appear from the singular points of 
$\hat \phi(x),$ however, multi-charge monopoles do not appear in general 
cases.


In this way, by the singular SU(2) gauge transformation,
there appears  the monopole with the Dirac string.
Here, we consider the role of the off-diagonal component in the SU(2)
gauge function $\Omega^H$ to appearance of the monopole, by comparing with the 
U(1)$_3$ gauge transformation.
Let us consider the singular gauge transformation
$\Omega^{\rm U(1)} = e^{i\varphi \tau_3} \in$ U(1)$_{3}$
instead of $ \Omega^H$. 
This U(1)$_{3}$ gauge function 
$\Omega^{\rm U(1)}$ is multi-valued on the whole region of the $z$ axis 
($\theta=0$ and $\theta=\pi$),
and  
$A_{\mu}^{\rm sing} \equiv -\frac{i}{e} \Omega^{\rm U(1)} 
\partial_{\mu} \Omega^{{\rm U(1)}\dagger}$ also has a singularity.
The magnetic flux which penetrates the area inside the 
closed 
contour $c(r,\theta) = \{ r,\theta,\varphi | 0 \le \varphi < 2 \pi \}$ 
is found to be
\begin{eqnarray}
\Phi^{\rm flux}( \theta ) 
=\int_c d{\bf x} \cdot {\bf A}^{\rm sing}   = - \frac{4\pi}{e}  \frac{\tau_3}{2},
\end{eqnarray}
which corresponds to the endless Dirac string along the $z$-axis.
It is noted that the singular U(1)$_3$ gauge transformation can provide 
the endless Dirac string, however, it never creates the monopole.

The monopole is created not by above singular U(1)$_3$ 
gauge transformation but by a singular SU(2) gauge transformation.
Since the multi-valuedness of $ \Omega^H$ is 
originated from 
the $\varphi$-dependence
at $\theta = 0$ or $\theta = \pi$, 
we separate the SU(2) gauge function  (\ref{eq:gauge-function}) as
\begin{eqnarray}
 \Omega = {\rm cos}\frac{\theta}{2} e^{i\varphi \tau_3}+  
( \varphi \mbox{-independent term} ). \nonumber
\end{eqnarray}
At $\theta=0$ or the positive side of $z$
 axis, $ \Omega^H $ coincides with  $\Omega^{\rm U(1)} \equiv e^{i 
 \varphi \tau_3}$ and is multi-valued like $\Omega^{\rm U(1)}$.
Therefore the Dirac string is created at $\theta = 0$ by the gauge
transformation $ \Omega^H$.
 On the other hand, at $\theta = \pi$ or the negative side of $z$-axis,
 $\varphi$-dependent part of $ \Omega$ vanishes due to $\cos 
 \frac{\theta}{2} = 0$, so that the Dirac string never appears in $ \Omega 
 \partial_\mu  \Omega^\dagger$ at $\theta = \pi$.
 Thus, by the SU(2) singular gauge transformation  $ \Omega^H$,
 the Dirac string is generated only on the positive side of the $z$-axis
 and terminates at the origin $r = 0$, and hence the monopole appears at the end 
 of the Dirac string.
Around the origin $ \tilde {\bf x}=0$, the factor $\cos \frac{\theta}{2}$ varies from 
unity to zero continuously 
with the polar angle $\theta$, and this makes the Dirac string 
terminated. 
Such a variation of the norm of the diagonal component  $\cos 
\frac{\theta}{2}
e^{i\varphi}$ cannot be realized in the U(1)$_3$ gauge transformation
with $\Omega^{\rm U(1)}$.
In the SU(2) gauge transformation with $\Omega^H$, the norm of the 
diagonal component can be changed owing to existence of the 
off-diagonal component of $\Omega^H$, and the difference of the 
multi-valuedness between $\theta=0$ and $\theta=\pi$ leads to the 
terminated Dirac string and the monopole. 
In this way, to create the monopole in QCD,
full SU(2) components of the (singular)
gauge transformation is necessary, and therefore one can expect a close 
relation between monopoles and the off-diagonal component of the gluon 
field.

\subsection{
Appearance of Monopole 
in the Gauge-connection Formalism}


In this subsection, we study the appearance of monopoles 
in the abelian sector of QCD in the abelian gauge in detail using
the gauge connection formalism.
In the abelian gauge, 
the monopole or the Dirac string appears as the result of 
the SU($N_{c}$) singular gauge transformation from a 
regular (continuous) gauge configuration. 
For the careful description of the singular gauge transformation, 
we formulate the gauge theory in terms of 
of the gauge connection, described 
by the covariant-derivative operator $\hat D_\mu$
and
$
\hat D_\mu  \equiv \hat \partial_\mu +ieA_\mu (x), 
$
where $\hat \partial_\mu $ is the derivative operator satisfying 
$[\hat \partial_\mu ,f(x)]=\partial_\mu f(x)$.

To begin with, let us consider the system holding the local 
difference of the internal-space coordinate frame. 
We attention the neighbor of the real space-time $x_\mu$,
and denote by $|q(x) \rangle$ the basis of the internal-coordinate frame.
At the neighboring point $x_\mu + \varepsilon_\mu$, we express the 
difference of the internal-coordinate frame as
$
|q(x+\varepsilon) \rangle  =
R_\varepsilon (x) |q(x) \rangle
$
with $R_\varepsilon (x) = e^{ir_\varepsilon(x)}$ $\in G$ being
the `rotational matrix' of the internal space.
We require the `local superposition' on $r_\varepsilon$ as
$r_{\varepsilon_1+\varepsilon_2}= r_{\varepsilon_1}+r_{\varepsilon_2}$
up to $O(\varepsilon)$, and then we can express 
$r_\varepsilon(x) = -e \varepsilon_\mu A^\mu(x)$ using a
$\varepsilon$-independent local variable $A_\mu(x)$ $\in$ $g$ :
$|q(x+\varepsilon) \rangle = e^{-i e\varepsilon_\mu A^\mu(x) } |q(x) \rangle.$
Then, the `observed difference' of the internal space 
coordinate depends on 
the real space-time $x_\mu $, 
the observed difference of the local operator $O(x)$ 
between neighboring points, 
$x_\mu $ and $x_\mu +\varepsilon _\mu$, is given by 
\begin{eqnarray}
&  & \langle q(x+\varepsilon )|O(x+\varepsilon )|q(x+\varepsilon ) \rangle  
 -  \langle q(x)|O(x)|q(x) \rangle  \nonumber \\
& = & 
 \langle q(x)| e^{ie \varepsilon _\mu  A^\mu (x)}O(x+\varepsilon ) 
 e^{-ie \varepsilon _\mu  A^\mu (x)}|q(x) \rangle - \langle q(x)|O(x)|q(x) 
\rangle
\nonumber \\
& \simeq & 
\varepsilon _\mu  \langle q(x)|\{\partial^\mu O(x)
 +ie[A^\mu (x), O(x)]\}|q(x) \rangle 
\nonumber \\
& = &
\varepsilon _\mu  \langle q(x)|\{[\hat \partial^\mu +ieA^\mu (x), 
O(x)]\}|q(x) \rangle 
\equiv
\varepsilon _\mu  \langle q(x)|[\hat D^\mu , O(x)]|q(x) \rangle. 
\end{eqnarray}
Here, one finds natural appearance of the covariant derivative operator, 
$
\hat D_\mu  \equiv \hat \partial_\mu +ieA_\mu (x).
$
The gauge transformation is simply defined by the 
arbitrary internal-space rotation as 
$|q(x) \rangle \rightarrow \Omega (x) |q(x) \rangle $
with $\Omega(x)$ $\in$ $G$,
and therefore the covariant derivative operator
is transformed as 
$
\hat D_\mu \rightarrow \hat D^\Omega_\mu  =\Omega (x)\hat D_\mu \Omega^\dagger 
(x)$
with $\Omega(x)$ $\in$ $G$,
which is consistent with 
$A_\mu \rightarrow A^\Omega_\mu =\Omega (A_\mu -\frac{i}{e}\partial_\mu )\Omega 
^\dagger$. 

In the general system including singularities such as the Dirac string,
the gauge field and the field strength are defined as
the difference between the gauge connection and the derivative 
connection,
\begin{eqnarray}
A_\mu & \equiv & \frac{1}{ie}({\hat D_\mu}-{\hat \partial_\mu}) \\
G_{\mu\nu} & \equiv & \frac{1}{ie}
([{\hat D_\mu},{\hat D_\nu}]-[\hat \partial_\mu,\hat \partial_\nu]).
\end{eqnarray}
This expression of $G_{\mu\nu}$ is returned to the standard definition 
$G_{\mu\nu} 
= \frac{1}{ie} [{\hat D_\mu},{\hat D_\nu}] 
= \partial_\mu A_\nu -\partial_\nu A_\mu + ie[A_\mu, A_\nu]$
in the regular system.
By the general gauge transformation with the gauge function $\Omega $, 
the field strength $G_{\mu\nu}$ is transformed as
\begin{eqnarray}
G_{\mu\nu} \rightarrow G^\Omega_{\mu\nu} & = &  
\Omega G_{\mu\nu} \Omega^{\dagger} 
 =  \frac{1}{ie} ([\hat D^\Omega_\mu ,\hat D^\Omega_\nu ]
-\Omega [\hat \partial_\mu , \hat \partial_\nu ]\Omega {}^\dagger) 
\nonumber \\
& = & \partial_\mu A^\Omega_\nu -\partial_\nu A^\Omega_\mu 
+ ie[A^\Omega_\mu, A^\Omega_\nu] +\frac{i}{e}
(\Omega[\hat \partial_\mu, \hat \partial_\nu] \Omega^{\dagger} 
-[\hat \partial_\mu, \hat \partial_\nu]) 
\nonumber \\
& = & (\partial_\mu A^\Omega_\nu -\partial_\nu A^\Omega_\mu )
+ ie[A^\Omega_\mu, A^\Omega_\nu] +\frac{i}{e}
\Omega[\partial_\mu, \partial_\nu] \Omega^{\dagger}
\nonumber \\ 
& \equiv  & G^{\rm linear}_{\mu\nu} 
+  G^{\rm bilinear}_{\mu\nu} + G^{\rm sing}_{\mu\nu}.
\label{eq:field-strength}
\end{eqnarray}
The last term remains only for the singular gauge transformation
on $\Omega^H$ and $\Omega^{\rm U(1)}$, 
and can provide the Dirac string.

Figure \ref{gfig7}
shows the SU(2) field 
strength $G_{\mu\nu}^{\rm linear}, 
G_{\mu\nu}^{\rm bilinear}$ and $ G_{\mu\nu}^{\rm sing}$ 
in the abelian gauge provided 
by $\Omega^H$ in Eq.(\ref{eq:gauge-function}).
The linear term $G^{\rm linear}_{\mu\nu} \equiv
(\partial_\mu A^\Omega_\nu -\partial_\nu A^\Omega_\mu ) $
includes in the abelian sector the singular gauge configuration of
{\it the monopole with the Dirac string}, which
supplies the magnetic flux from infinity.
Since each component satisfies the Bianchi identity 
$\partial^\alpha {}^* G^{ \rm linear}_{\alpha\mu} = \partial^\alpha {}^*
(\partial \wedge A^\Omega )_{\alpha\mu}=0$,
the abelian magnetic flux is conserved.
The abelian part of $ G^{\rm bilinear}_{\mu\nu}
\equiv ie[A^\Omega_\mu, A^\Omega_\nu]$, $ (G^{\rm bilinear}_{\mu\nu})^3
= -e(A_{\mu}^1 A_{\nu}^2-  A_{\nu}^1 A_{\mu}^2)$, 
includes the effect  
of off-diagonal components, and it is dropped by the abelian projection.
The last term
$G^{\rm sing}_{\mu\nu} \equiv 
\frac{i}{e}
\Omega[\partial_\mu, \partial_\nu]  \Omega^{\dagger}$
appears from the singularity of the gauge function $ \Omega$, and 
it plays the important role of the appearance of the magnetic monopole 
in the abelian sector.

First, we consider the singular part $G^{\rm sing}_{\mu\nu}$.
In general, $G^{\rm sing}_{\mu\nu}$ disappears in the regular point 
in $\Omega$. 
It is to be noted that $G^{\rm sing}_{\mu\nu}$ is found to be diagonal 
from the direct calculation  with  $\Omega^H$ in Eq.(\ref{eq:gauge-function}),
\begin{eqnarray}
G^{\rm sing}_{\mu\nu}  & \equiv & 
\frac{i}{e} \Omega^H[\partial_\mu, \partial_\nu] \Omega^{H\dagger} \nonumber 
 =   \frac{i}{e}(g_{\mu1} g_{\nu2} - g_{\mu2} g_{\nu1})
{\rm cos} ^2 \frac{\theta}{2} e^{i \varphi \tau_3} [\partial_1, \partial_2]
e^{-i \varphi \tau_3} \nonumber  \\
& = & \frac{1}{e} (g_{\mu1} g_{\nu2} - g_{\mu2} g_{\nu1})
\frac{1+ {\rm cos} \theta}{2} [\partial_1, \partial_2]  \varphi
\cdot \tau_3  
\nonumber \\ & = &
\frac{4 \pi}{e} (g_{\mu1} g_{\nu2} - g_{\mu2} g_{\nu1}
)
\theta(x_3)  \delta(x_1) \delta(x_2) \cdot \frac{\tau_3}{2},
\label{eq:monopole}
\end{eqnarray}
where we have used relations,
\begin{eqnarray}
[\partial_1, \partial_2]\varphi = - 2 \pi \delta(x_1) \delta(x_2), 
\hspace{0.4cm}
 \frac{1+ {\rm cos} \theta}{2} \delta(x_1) 
\delta(x_2)
= \theta(x_3) \delta(x_1) \delta(x_2). 
\end{eqnarray}
The off-diagonal component of $ \Omega ^H [\partial_\mu, \partial_\nu] 
\Omega^{H\dagger}$ disappears, since the 
singularity appears only from $\varphi$-dependent term. 
As a remarkable fact,
the last expression  in Eq.(\ref{eq:monopole})  
shows the terminated Dirac string,
which is placed along the positive $z$-axis with the end at the origin.
Hence, in the abelian part of the SU(2) field strength,
$G^{\rm sing}_{\mu\nu}$ leads to the breaking of the U(1)$_3$ Bianchi identity,
\begin{eqnarray}
k_\mu  & = & \partial^\alpha {}^* G^{\rm sing}_{\alpha\mu} 
= \frac12  \varepsilon_{\alpha\mu}{}^{\beta\gamma}
\partial^\alpha G^{\rm sing}_{\beta\gamma}
= \frac{4\pi}{e}  
 \varepsilon_{\alpha\mu12}\partial^\alpha \{ \delta(x_1)\delta(x_2)\theta(x_3) 
\}
\frac{\tau_3}{2}  \nonumber \\
 & = & 
 \frac{4\pi}{e} g_{\mu0} \delta(x_1)\delta(x_2)\delta(x_3) 
\frac{\tau_3}{2},
\label{eq:kk}
\end{eqnarray}
which is the expression for the static monopole
with the magnetic charge $g = \frac{4\pi}{e}$  at the origin.
Thus, the magnetic current $k_\mu$ is induced in the abelian sector
by the singular gauge transformation with $\Omega^H$ and {\it the Dirac 
condition $eg = 4 \pi$ is automatically derived} in this 
gauge-connection formalism.


In the covariant manner, 
$G^{\rm sing}_{\mu\nu}$ is expressed as
$G^{\rm sing}_{\mu\nu} = \frac{1}{n \cdot \partial } {}^{*}(n \wedge 
k)_{\mu\nu}$ 
using the monopole current $k_\mu$ in Eq.(\ref{eq:kk}) and a
constant 
4-vector $n_\mu$.
Actually, for the above case, one finds for $n_{\mu} = g_{\mu3}$
\begin{eqnarray}
\frac{1}{n \cdot \partial} {}^{*}(n \wedge k)_{\mu\nu}
\label{eq:k}
& = & 
\int dx^{'}_3 \langle x_3|\frac{1}{n \cdot \partial}|x^{'}_3 \rangle 
\varepsilon_{\mu\nu30} n^3
\frac{4\pi}{e} \delta (x_1) \delta(x_2) \delta(x^{'}_3) \frac{\tau_3}{2} 
\nonumber  \\
& = & \frac{4 \pi}{e} (g_{\mu1} g_{\nu2} -
g_{\mu2} g_{\nu1}) \theta(x_3)
\delta(x_1) \delta(x_2)  \frac{\tau_3}{2} \nonumber \\ 
& = & 
\frac{i}{e}  \Omega^H[\partial_\mu, \partial_\nu]  \Omega^{H\dagger}
= G^{\rm sing}_{\mu\nu},
\end{eqnarray}
using the relation 
$ \langle x_n|\frac{1}{n \cdot \partial}|x_n^{'} \rangle 
 = \theta(x_n-x_n^{'})$.  

Thus, the last term $G^{\rm sing}_{\mu\nu}$ 
corresponds to the Dirac string terminated at the origin.
Since $G^{\rm linear}_{\mu\nu}$  shows 
the configuration of the monopole together with the Dirac string,
the sum of $G^{\rm linear}_{\mu\nu} + G^{\rm sing}_{\mu\nu}$
provides the gauge configuration of the monopole
without the Dirac string in the abelian sector.
Thus, 
by dropping the off-diagonal gluon element,
$G^{\rm bilinear}_{\mu\nu}$ vanishes and 
the remaining part $(G_{\mu\nu}^{\rm linear} +G_{\mu\nu}^{\rm sing})^3$
describing the abelian projected QCD includes the field 
strength of monopoles.

Next, we consider the role of off-diagonal gluon components for  
appearance of the monopole.
The gluon  field 
 is divided into the regular part 
$ \Omega  A_\mu  \Omega^{\dagger}$
and the singular part
$   -\frac{i}{e}  \Omega  \partial_\mu  \Omega^{\dagger}$.
Since we are interested in the behavior of the singularity,
we neglect the regular part of the gluon field.
Then, $G^{\rm bilinear}_{\mu\nu}$ is written as
\begin{eqnarray}
ie[A^\Omega_\mu, A^\Omega_\nu]  & = &
\frac{1}{ie}[ \Omega \partial_\mu  \Omega^{\dagger},
              \Omega \partial_\nu  \Omega^{\dagger}]  \nonumber  \\
& = & - \frac{1}{ie}\{ (\partial_\mu  \Omega) \partial_\nu  
\Omega^{\dagger}
                 - (\partial_\nu  \Omega) \partial_\mu  \Omega^{\dagger} 
\} 
                 \nonumber \\
& = & -\frac{1}{ie} \{ \partial_\mu ( \Omega  \partial_\nu  
\Omega^{\dagger} )
                 - \partial_\nu ( \Omega  \partial_\mu  \Omega^{\dagger} ) 
\}
  -\frac{i}{e}    \Omega[\partial_\mu, \partial_\nu]  \Omega^{\dagger} 
  \nonumber \\
& = & -(\partial_\mu A^\Omega_\nu -\partial_\nu A^\Omega_\mu) 
 -\frac{i}{e}    \Omega[\partial_\mu, \partial_\nu]  \Omega^{\dagger}, 
\end{eqnarray}
where the last term appears as the breaking of the Maurer-Cartan 
equation.
In the abelian gauge, the singularity 
of the monopole  appearing in  
$G^{\rm linear}_{\mu\nu} + G^{\rm sing}_{\mu\nu}$ is exactly canceled by  that of  
$G^{\rm bilinear}_{\mu\nu}$. 
Thus, in the abelian gauge,  the off-diagonal gluon combination 
$(G_{\mu\nu}^{\rm    bilinear})^3 = -e
\{ \, (A_{\mu}^{\Omega})^1 \, (A_{\nu}^{\Omega})^2 - 
(A_{\nu}^{\Omega})^1  (A_{\mu}^{\Omega})^2 \, \}$ includes the field 
strength of the anti-monopole, and hence
the off-diagonal gluons $(A_{\mu}^{\Omega})^1$ and $(A_{\mu}^{\Omega})^2$
 have to include some singular structure around the monopole.

The abelian projection is defined by
dropping the off-diagonal component of the gluon field $A_\mu$,
\begin{eqnarray}
A_\mu^\Omega \equiv A_{\mu a}^\Omega \frac{\tau^a}{2} \rightarrow 
{\cal A}_\mu \equiv {\rm tr}(A_\mu^\Omega \tau^3) \frac{\tau^3}{2} = 
 (A_\mu^\Omega)^3 \frac{\tau^3}{2}.
\end{eqnarray}
Accordingly, the SU(2) field strength $G_{\mu\nu}^\Omega$ is projected to 
the abelian field strength 
${\cal F}_{\mu\nu} \equiv F_{\mu\nu}  \frac{\tau^3}{2}$,
\begin{eqnarray}
G_{\mu\nu}^\Omega  & \equiv & (G_{\mu\nu}^\Omega)^a \frac{\tau^a}{2}
 =  (\partial_\mu A^\Omega_\nu -\partial_\nu A^\Omega_\mu )
+ ie[A^\Omega_\mu, A^\Omega_\nu] +\frac{i}{e}
\Omega[\partial_\mu, \partial_\nu] \Omega^{\dagger} \nonumber \\
&  \rightarrow   &{\cal F}_{\mu\nu}  = 
\partial_\mu {\cal A}_\nu -\partial_\nu {\cal A}_\mu +\frac{i}{e}
 \Omega[\partial_\mu, \partial_\nu]  \Omega^{\dagger} \nonumber \\
& &  = 
\partial_\mu {\cal A}_\nu -\partial_\nu {\cal A}_\mu - 
{\cal F}_{\mu\nu}^{\rm sing},
\end{eqnarray}
where ${\cal F}_{\mu\nu}^{\rm sing} \equiv  
{F}_{\mu\nu}^{\rm sing}  \frac{\tau_3}{2}  \equiv -
\frac{i}{e} \Omega[\partial_\mu, \partial_\nu]  \Omega^{\dagger}  $ is diagonal
and remains.
Here, the bilinear term 
 $ie[A^\Omega_\mu, A^\Omega_\nu]$ vanishes in  AP-QCD because 
 it is projected to $ie[{\cal A}_\mu, {\cal A}_\nu]=0$ by the abelian 
 projection.
The appearance of ${\cal F}_{\mu\nu}^{\rm sing} $ 
 leads to the breaking of the abelian  Bianchi identity in the U(1)$_3$ sector,
\begin{eqnarray}
\partial^\alpha {}^*{\cal F}_{\alpha\mu}  =
- \partial^\alpha {}^*{\cal F}_{\alpha\mu}^{\rm sing} 
= \partial^\alpha {}^*  \{ \frac{i}{e} \Omega[\partial_\alpha, 
    \partial_\mu]  \Omega^{\dagger} \}
= k_\mu,
\end{eqnarray}
where Eq.(\ref{eq:k}) is used.
Thus, the magnetic current $k_\mu$ is induced into the abelian gauge theory 
through the singularity of the SU(2)  gauge transformation. 

Here, we compare AP-QCD and QCD in terms of the field strength.
The SU($N_{c}$) field strength $G_{\mu\nu}$ is controlled by the QCD action,
$
S_{\rm QCD} = \int d^4x \{ -\frac12 {\rm tr}G_{\mu\nu}G^{\mu\nu}\},
$
so that each component $G_{\mu\nu}^a$ cannot diverge.
On the other hand, the field strength 
${\cal F}_{\mu\nu}$ in AP-QCD is not directly
controlled by $S_{\rm QCD}$, since the QCD action includes
also off-diagonal components.  
It should be noted that the point-like monopole appearing in AP-QCD 
makes the  U(1)$_3$ action 
$
S_{\rm Abel} = \int d^4x \{ -\frac12 {\rm tr}{\cal F}_{\mu\nu}{\cal 
F}^{\mu\nu} \}
$
divergent around the monopole, 
such a divergence  in ${\cal F}$ should cancel exactly with the 
remaining off-diagonal contribution from 
$G^{\rm bilinear}_{\mu\nu}$ to keep the total QCD action finite.
Thus, the appearance of monopoles in AP-QCD is supported by the 
singular contribution of off-diagonal gluons.
In this way, abelian projected QCD includes monopoles generally.


\subsection{Monopole Current in the Lattice Formalism}

For the study of nonperturbative QCD physics, the lattice QCD 
formalism provides a useful method for the direct calculation 
of the QCD generating functional $Z_{\rm QCD}$\cite{rothe}. 
In this subsection, we extract the abelian gauge field and
the monopole current in the lattice formalism \cite{degrand}.

In the lattice QCD, 
the system is described by the link-variable $U_\mu(s) \equiv e^{iaeA_\mu(s)}$
$\in$ SU$(N_c)$ instead of $A_\mu(x)$. Here, $e$ denotes the QCD gauge 
coupling and $a$ the lattice spacing. 
The SU(2) link-variable $U_\mu(s)$ can be 
factorized as 
\begin{eqnarray}
U_\mu(s) & = & 
M_\mu(s)  u_\mu(s)   \hspace{3cm} \in G \nonumber \\
M_\mu(s) & = & 
{\rm exp} \left(
i \{  \tau_1 \theta^1_\mu(s)+ \tau_2 \theta^2_\mu(s)  \} 
\right) \hspace{1cm} 
\in  G/H, \nonumber \\
u_\mu(s) & = & {\rm exp} \left( {i \tau^3 \theta^3_\mu (s) } \right) 
\hspace{3cm} \in  H
\label{eq:para}
\end{eqnarray}
with respect to the Cartan decomposition of $G = G/H$ $\times$ $H$ into 
$G/H=$SU(2)/U(1)$_3$ 
and $H=$U(1)$_3$.
In the lattice formalism, such a factorization has an ambiguity relating 
to the ordering of $M_\mu$ and  $u_\mu$ in this factorization.
Instead of $U_\mu = M_\mu u_\mu$, another facotrization 
 $U_\mu = u_\mu M_\mu$ is equivalently applicable, however, 
 such an ordering is to be fixed through the whole argument.
Here, the abelian link variable, 
\begin{eqnarray}
u_\mu(s) = e^{i \tau^3  \theta^3_\mu (s) } =
 \left( {\matrix{
e^{i\theta^3_\mu  (s)}  &          0
\cr
0                     &  e^{-i\theta^3_\mu (s)} \cr
}} \right)  \hspace{1cm}  \in {\rm U(1)}_3 \subset {\rm SU(2)}, 
\end{eqnarray}
plays the similar role as the SU(2) link-variable 
$U_\mu(s) \in {\rm SU(2)}$ 
in terms of the residual U(1)$_3$ gauge symmetry 
in the abelian gauge, and $\theta^3_\mu(s) \in (-\pi, \pi]$ 
corresponds to the diagonal component of the gluon 
in the continuum limit. 
On the other hand, the off-diagonal factor 
$M_\mu(s) \in $ SU(2)/U(1)$_3$ is expressed as 
\begin{eqnarray}
 M_\mu(s)  & = & 
{\rm exp} \left(
{i \{  \tau_1 \theta^1_\mu(s)+ \tau_2 \theta^2_\mu(s)  \} }
\right)
\nonumber \\
& = & 
\left( 
{\matrix{
{\rm cos}{\theta_\mu(s)} & -{\rm sin}{\theta_\mu(s)}e^{-i\chi_\mu(s)} \cr
{\rm sin}{\theta_\mu(s)}e^{i\chi_\mu(s)} & {\rm cos}{\theta_\mu(s)} 
}}
 \right) 
\label{divided} 
 \\ & = &
\left( {\matrix{
\sqrt{1-|c_\mu(s)|^2} & -c_\mu^*(s) \cr
c_\mu (s) & \sqrt{1-|c_\mu(s)|^2}
}} \right)  \nonumber 
\end{eqnarray}
with  
$\theta_\mu (s) \equiv {\rm mod}_\frac{\pi}{2} \sqrt{ (\theta^1_\mu )^2 + 
(\theta^2_\mu)^2} \in [0, \frac{\pi}{2}]$ and $\chi_\mu (s) \in ( - \pi, \pi]$.
Near the continuum limit, the off-diagonal elements of $M_\mu(s)$ 
correspond to the off-diagonal gluon components. 
Under 
the residual U(1)$_3$ gauge transformation by 
$\omega(s) = e^{-i \varphi(s) \frac{\tau_3}{2}} \in$ U(1)$_3$, 
$u_\mu(s)$ and $M_\mu(s)$ are transformed as
\begin{eqnarray}
u_\mu(s) &\rightarrow & 
u^\omega _\mu(s) = \omega(s) u_\mu(s) \omega^\dagger(s+ \hat \mu 
)  \hspace{1cm}  \in H\\
M_\mu(s) &\rightarrow &
M^\omega_\mu(s) = \omega(s) M_\mu(s) \omega^\dagger(s) \hspace{1cm} \in G/H
\end{eqnarray}
so as to keep $M_\mu^\omega(s)$ belong $G/H$.
Accordingly, 
$\theta^3_\mu(s)$ and $c_\mu(s) \in {\bf C}$ are transformed as 
\begin{eqnarray}
\theta^3_\mu(s) &\rightarrow & 
\theta^{3 \omega}_\mu(s)=
{\rm mod}_{2\pi} [ \theta^3_\mu(s) + \{ \varphi(s+ \hat \mu)-\varphi(s) \} /2 ]
\\
c_\mu(s) &\rightarrow & 
c_\mu^{\omega}(s) = c_\mu(s)e^{i\varphi(s)}.
\end{eqnarray}
Thus, on the residual U(1)$_3$ gauge symmetry, 
$u_\mu(s)$ behaves as the U(1)$_3$ lattice gauge field, and 
$\theta^3_\mu(s)$ behaves as the U(1)$_3$ gauge field 
in the continuum limit. 
On the other hand, 
$M_\mu(s)$ and $c_\mu(s)$ behave as the charged matter field 
in terms of the residual U(1)$_3$ gauge symmetry, which is 
similar to the charged weak boson $W_\mu^{\pm}$ in the Standard Model.

The abelian field strength ${\bar \theta_{\mu\nu}}(s)$ is defined as 
$
{\bar \theta_{\mu\nu}} (s) \equiv {\rm mod }_{2\pi} 
 (\partial \wedge \theta^3)_{\mu\nu}(s) \in (-\pi,\pi]$,
which is U$(1)_{3}$ gauge invariant.
In general, the two form of the abelian angle variable 
 $\theta^3_\mu(s) $ is divided as
\begin{eqnarray}
\theta_{\mu\nu}(s) \equiv (\partial \wedge \theta^3)_{\mu\nu}(s)
= {\bar \theta_{\mu\nu} }(s) + 2 \pi n_{\mu\nu}(s),
\label{eq:twoform}
\end{eqnarray}
where $n_{\mu\nu}(s) \in {\bf Z}$
corresponds to the quantized magnetic flux of 
the `Dirac string' penetrating through the plaquette.  
Although $n_{\mu\nu} \ne 0$ provides the infinite magnetic field is 
the continuum limit as $2 \pi n_{\mu\nu}/a$,
the term $2 \pi n_{\mu\nu}(s)$ does not contribute to the abelian
plaquette $\Box^{\rm Abel}_{\mu\nu}(s)$,  
and it is changed by the singular U(1)$_3$ gauge-transformation as
$\theta^3_\mu(s) \rightarrow \theta^3_\mu(s)+ \partial_\mu \varphi(s) $
with $\varphi(s)$ being the azimuthal angle.
Thus,  $2 \pi n_{\mu\nu}$ corresponds to the Dirac string as an 
unphysical object.

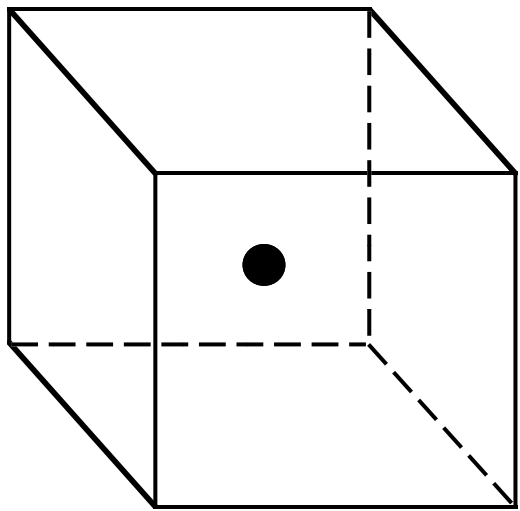
\begin{figure}[bt]
\caption{The (static) monopole defined on the dual lattice
is equivalent to the total magnetic flux of the Dirac string,
and therefore the magnetic charge of the QCD-monopole is quantized.
}
\label{monopole}
\end{figure}

The monopole $k^{lat}_\mu(s)$ is defined on the {\it dual link} as
\cite{degrand},
\begin{eqnarray}
k^{lat}_\mu(s)  \equiv  \frac{1}{2\pi} \partial_\alpha {}^*{\bar 
\theta_{\alpha\mu}} (s)
 = - \partial_\alpha  {}^*n_{\alpha\mu}(s),
 \label{eq:monodif}
\end{eqnarray}
using the abelian field strength $\bar \theta_{\mu\nu}(s)$.
Here,  $k^{lat}_\mu(s) $ is 
defined such that the topological quantization is manifest,
 $k^{lat}_\mu(s) \in {\bf Z}$.
In this definition,  
for instance, one finds    
$k^{lat}_0 =  \frac12 \varepsilon_{ijk} \partial_i n_{jk}$ and
$k^{lat}_i = 0$ ($i=1,2,3$) for the static monopole.
The magnetic charge of the monopole on the dual lattice is
determined
by the total magnetic flux of the Dirac strings
entering
the cube around the monopole. (See Fig.\ref{monopole}.)

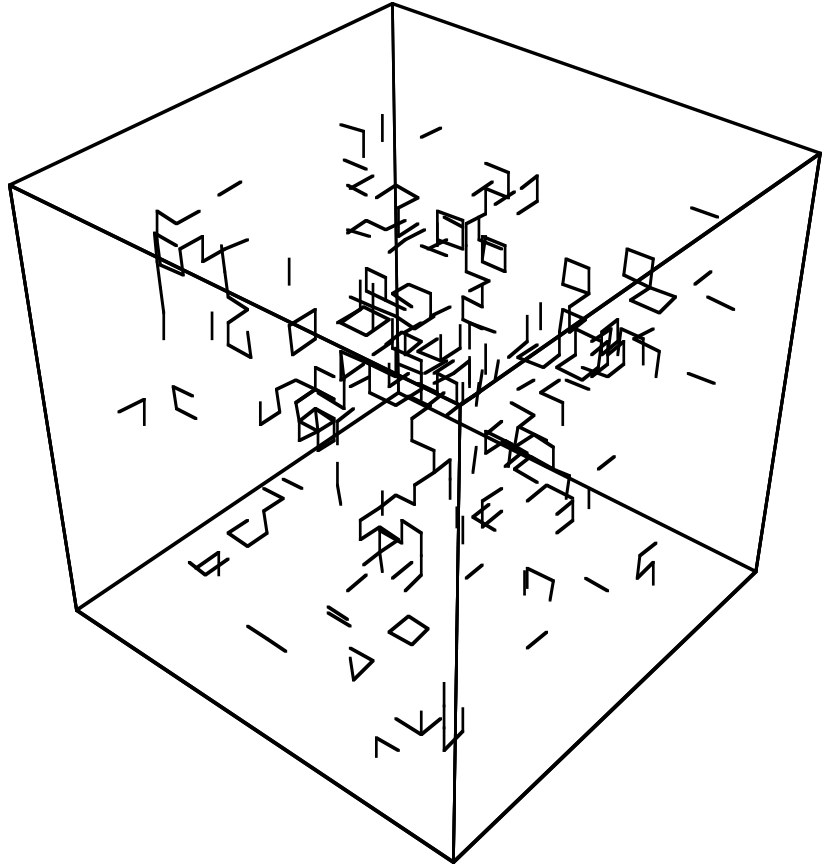
\begin{figure}[bt]
\caption{The typical example of the 3-dimensional time-slice of the 
monopole current in the MA gauge  in the lattice QCD with 
$\beta$ = 2.4 on $16^4$.
}
\label{monopolecur}
\vspace{0cm}
\end{figure}

We show in Fig.\ref{monopolecur} the typical example of the
monopole current at a time slice in the lattice QCD
at $\beta$ = 2.4 in the maximally abelian (MA) gauge.
In each gauge configuration, the monopole current 
appears as a distinct line-like object, and 
the neighbor of the monopole can be defined 
on the lattice. However, taking the temporal direction into 
account, the monopole current forms a global network covering over 
${\bf R}^4$.
 
Here, we summarize several relevant properties of $k_\mu(s)$.
\begin{enumerate}
\item
The monopole current $k_\mu$ is topologically quantized 
and $k^{lat}_\mu(s)$ takes an integer 
$k^{lat}_\mu(s)$ $\in$ ${\bf Z}$ in the definition of 
Eq.(\ref{eq:monodif}).
As the result, 
$k^{lat}_\mu(s)$ forms 
a line-like object in the space-time ${\bf R}^4$,
since $k^{lat}_\mu$ is a conserved current as $\partial_\mu k_\mu = 0$.
These features of $k_\mu^{lat}(s)$ $\in$ ${\bf Z}$ are quite unique and 
different from the electric current $j_\mu(s)$ $\in$ ${\bf R}$, which can 
spread as a continuous field.  
\item
In the lattice formalism, 
$k^{lat}_\mu \equiv \frac{1}{2\pi} \partial_\alpha ^* \bar 
\theta_{\alpha\mu}$ 
is defined as a three-form on the dual link. 
For the use of the forward derivative, $k^{lat}_\mu(s)$ is to be 
defined on the dual link between 
$s^{\rm dual}_{\pm \mu} \equiv 
s+\frac{\hat x}{2}+\frac{\hat y}{2}+\frac{\hat z}{2}+\frac{\hat t}{2}
\pm \frac{\hat \mu}{2}$.
For instance, $k^{lat}_0(s)$ is placed on the dual link between 
$(s_x+\frac12,s_y+\frac12,s_z+\frac12,s_t)$ and 
$(s_x+\frac12,s_y+\frac12,s_z+\frac12,s_t+1)$.
Thus, the monopole is defined to appear 
at the center of the 3-dimensional cube 
perpendicular to the monopole-current direction 
as shown in Fig.\ref{monopole}.
\item
Because of 
$k_\mu \equiv \partial_\alpha ^* F_{\alpha\mu}
=-\frac{1}{2} \varepsilon_{\mu\alpha\beta\gamma}
\partial_\alpha F_{\beta\gamma}$, 
$k_\mu$ only affects the perpendicular components to the 
$\hat \mu$-direction
for the `electric variable' as $F_{\alpha\beta}$
 in a direct manner.
For instance, the static monopole with $k_0 \ne 0$ creates 
the magnetic field $F_{ij}$ ($i,j$=1,2,3) around it, but 
does not bring the electric field $F_{0i}$.
Hence, in testing the field around the monopole, 
one has to consider the difference between 
such perpendicular components and others. 
\end{enumerate}

We now consider the relationship between the lattice variable
and the field variable in the continuum theory.
The continuous abelian field 
${\cal A}_\mu (x)\equiv A^3_\mu (x) \frac{\tau^3}{2}$ is expressed as 
\begin{eqnarray}
eA^3_\mu \equiv  \theta^3_{\mu} \cdot \frac{2}{a}
\end{eqnarray}
with the gauge coupling constant $e$ and the lattice spacing $a$.
The abelian field strength
${\cal F}_{\mu\nu} (x)\equiv F_{\mu\nu} (x) \frac{\tau^3}{2}$ 
 in the continuum theory is written as
\begin{eqnarray}
eF_{\mu\nu} & \equiv  &  {\rm mod}_{2 \pi} (\theta_{\mu\nu}) 
\cdot \frac{2}{a^2}
=  {\bar \theta_{\mu\nu}} \cdot \frac{2}{a^2}, 
\end{eqnarray}
and $F_{\mu\nu}$ is composed of 
two parts according to the decomposition (\ref{eq:twoform})
\begin{eqnarray}
F_{\mu\nu} & = & (\partial \wedge A^3)_{\mu\nu} - F^{\rm sing}_{\mu\nu}.
\label{eq:twopars}
\end{eqnarray}
Thus, in the SU$(N_c)$-lattice formalism, the difference between the 
field strength $F_{\mu\nu}$ and two-form $(\partial \wedge A)_{\mu\nu}$ 
arises from the periodicity of the angle variable in the compact subgroup 
U(1)$^{N_c-1}$ embedded in SU($N_c$).
Here, the singular Dirac-string part $F^{\rm sing}_{\mu\nu}$ is
directly related to $2 \pi n_{\mu\nu}$ and is 
written by
\begin{eqnarray}
F^{\rm sing}_{\mu\nu} =  2\pi n_{\mu\nu} \cdot \frac{2}{ea^2} = 
\frac{4\pi}{e} 
n_{\mu\nu}  \frac{1}{a^2}.
\end{eqnarray}
Owing to existence of $F^{\rm sing}_{\mu\nu}$ in Eq.(\ref{eq:twopars}),
the monopole current 
$k_{\mu} (x) \equiv k^3_{\mu} (x) \frac{\tau^3}{2}
\equiv \partial_\alpha {}^* F_{\alpha\mu} \frac{\tau^3}{2}$ 
appears in the continuum theory and is written as
\begin{eqnarray}
k^3_{\mu}  =
  k^{lat}_{\mu} \cdot \frac{4\pi}{ea^3}
= -\frac{4\pi}{e}  \partial_\alpha {}^*  n_{\alpha\mu} \frac{1}{a^3},
\end{eqnarray}
where the magnetic-charge unit $g \equiv \frac{4\pi}{e}$
naturally appears in $k_\mu$.

In the lattice formalism, there also appears the monopole-like 
configuration 
as the lattice artifact, when the lattice constant $a$ is relatively large.
As $a \rightarrow 0$, such a monopole-like configuration turns out to be 
regular large fluctuation rather than the point-like singularity.
Hence, one should use a fine mesh lattice to remove such lattice 
artifact monopoles.

\section{Monopoles in the Maximally Abelian Gauge}

\subsection{Maximally Abelian Gauge and Abelian Projection Rate}

The abelian gauge has some arbitrariness
corresponding to the choice of the variable $\phi[A_\mu(x)]$
to be diagonalized. 
Several typical abelian gauges
have been tested on the dual superconductor scenario for the 
nonperturbative QCD\cite{diacomo,poly}.
Recent lattice QCD studies show that 
infrared phenomena such as confinement properties and chiral symmetry
breaking are almost reproduced only by the abelian variable in the 
maximally abelian (MA) 
gauge \cite{ichiead,yotsuyanagi,hioki,bali,miyamura,woloshyn}. 
In this subsection, we study the MA gauge in detail, considering the 
gluon field properties.

In the SU(2) lattice formalism, the MA gauge is defined so as to maximize
\begin{eqnarray}
R_{\rm MA}[U_\mu]  & \equiv &
\sum_{s,\mu} {\rm tr} \{  U_\mu(s) \tau_3 U^{\dagger}_\mu(s) \tau_3 \}  
\nonumber \\
   & = & 
2\sum_{s,\mu}\{
U^0_\mu(s)^2+U^3_\mu(s)^2-U^1_\mu(s)^2-U^2_\mu(s)^2 \} \nonumber \\
   & = & 
2\sum_{s,\mu} \left[
1 - 2\{ U^1_\mu(s)^2+U^2_\mu(s)^2 \} \right]
\end{eqnarray}
by the SU(2) gauge transformation.
Here, we denote $U_{\mu}(s) \equiv U_{\mu}^0(s) + i \tau^a U_{\mu}^a (s)$
with $U_{\mu}^0(s),$ $U_{\mu}^a(s)$ $\in$ ${\bf R}$, obeying 
$U_{\mu}^0(s)^2 +  U_{\mu}^a(s)^2 = 1$.


The MA gauge is a sort of the abelian gauge which diagonalizes the 
hermite variable
\begin{eqnarray}
\Phi[U_{\mu}(s)] \equiv \sum _{\mu,\pm } U_{\pm \mu}(s) \tau_3 U^{\dagger}_{\pm 
\mu}(s).
\label{eq:operator}
\end{eqnarray}
Here, we use the convenient notation $U_{-\mu}(s) \equiv
U^{\dagger}_\mu(s- \hat \mu)$ in this paper.
Here, $\Phi[U_\mu(s)]$ is gauge transformed by $V(s)$ $\in$ SU(2) as
\begin{eqnarray}
\Phi(s) \rightarrow \Phi^V(s) = V(s)
\{  \sum_{\mu, \pm} U_\pm(s)V^\dagger(s\pm \hat \mu) \tau_3 V(s \pm \hat \mu)
U^{-1}_{\pm\mu}(s) \} V^\dagger(s),
\end{eqnarray}
which is not a simple adjoint transformation.
In the continuum limit $a \rightarrow 0$,
the link-variable reads $U_\mu(s) = e^{iaeA_\mu(s)} =1 + iaeA_\mu(s) 
+O(a^2)$, and hence
the MA gauge condition becomes 
$\displaystyle \sum_\mu$$(i\partial_\mu \pm e A^3_{\mu} ) A^{\pm }_\mu = 
0$, which can be regarded as the maximal decoupling condition 
between the abelian gauge sector and the charged gluon sector. 

In the MA gauge, $ \Phi(s)$ is diagonalized as 
$\Phi_{\rm diag}(s) =  \lambda(s) \frac{\tau_{3}}{2}$ with  $ 
\lambda(s) \in$ 
{\bf R}, and there remain the local U(1)$_3$ symmetry and the  global Weyl 
symmetry
\cite{suganuma4}.
After the MA gauge fixing, the global Weyl transformation with $W$ in 
Eq.(\ref{eq:weyl}) never changes
the sign of $\lambda(s)$ as 
\begin{eqnarray}
\Phi_{\rm diag}(s) \rightarrow 
\Phi_{\rm diag}^W (s) & = &  
\sum _{\mu,\pm }  W  U_{\pm \mu}(s)  W ^{\dagger} \tau_3  W  U^{\dagger}_{\pm \mu}(s)
 W^{\dagger} \\
& = &  
- \sum _{\mu,\pm } W  U_{\pm \mu}(s)   
\tau_3  U^{\dagger}_{\pm \mu}(s)
 W ^{\dagger}
= - W\Phi_{\rm diag}(s) W^\dagger  = \Phi_{\rm diag}(s).  \nonumber
\end{eqnarray}
Thus, the Weyl symmetry is not fixed in the MA gauge by the simple 
ordering condition as $\lambda(s) \ge 0$, 
unlike the simple adjoint case.  

In the MA gauge, the absolute values of the off-diagonal components, 
$U_\mu^1(s)$ and $U_\mu^2(s)$, are
forced to be small.
In  the continuum limit $a \rightarrow 0$,
the MA gauge is found to minimize the functional
\begin{eqnarray}
R_{ch}[A_\mu] \equiv \frac12 e^2 \int d^4 x \{ A_\mu^1(x)^2 + 
A_\mu^2(x)^2 \}
= e^2 \int d^4 x A_\mu^+(x) A_\mu^-(x)
\end{eqnarray}
with $A_\mu^\pm (x) \equiv {1 \over {\sqrt 2}}
  \{ A_\mu^1(x) \pm 
i A_\mu^2(x) \}$.
Thus, in the MA gauge, the off-diagonal gluon component is globally forced 
to be small by the gauge transformation, 
which seems a microscopic origin of abelian dominance for the 
nonperturbative QCD in the MA gauge
\cite{ichiead}.


Here, let us consider resemblance of the abelian link variable 
$u_{\mu}(s)$ to the SU(2) link variable $U_{\mu}(s)$ quantitatively.
To this end, 
we introduce the `abelian projection rate' $R_{\rm {Abel}}$ 
\cite{ichiead,poulis},
which is 
defined as the overlapping factor as 
\begin{eqnarray}
R_{\rm Abel} (s,\mu)  & \equiv  & \frac12 {\rm Re} \; {\rm tr} \{ U_{\mu}(s) 
u_{\mu}^{\dagger}(s) \} \nonumber \\ & = & 
\frac12 {\rm Re} \; {\rm tr} M_{\mu}(s) = \cos 
\theta_\mu(s)  \in[0,1], 
\end{eqnarray}
where $\theta_{\mu}(s)$ is defined to belong $[0,\frac{\pi}{2}]$ in 
the decomposition of $U_{\mu}(s)$ into $M_{\mu}(s)$ and $u_{\mu}(s)$. 
For instance, 
the SU(2) link variable $U_\mu(s)$ becomes completely abelian for $R_{\rm 
Abel}(s,\mu) = 1$,
while  $U_\mu(s)$ becomes completely off-diagonal for $R_{\rm 
Abel}(s,\mu) = 0$.
This definition of $R_{\rm {Abel}}$ is inspired by the ordinary 
`distance' between two 
matrices $A, B \in {\rm GL}(N,{\bf C})$ defined as
$d^{2}(A,B) \equiv \frac12 {\rm tr} \{ (A-B)^{\dagger} (A-B) 
\}$\cite{georgi}, 
which 
leads to 
$d^{2}(A,B) = 2-{\rm Re} \;{\rm tr}(AB^{\dagger})$ for $A,B \in$SU(2).
In fact, the similarity between $U_\mu(s)$ and $u_\mu(s)$ 
can be quantitatively measured 
in terms of the `distance' between them.
In the strong-coupling limit ($\beta = 0$),
$\langle  R_{\rm Abel} \rangle_{\beta=0} = \langle  \cos \theta_\mu(s) 
\rangle _{\beta=0} $ without gauge fixing 
is analytically 
calculable as  \cite{ichie6,poulis}
\begin{eqnarray}
\langle  R_{\rm Abel}(s,\mu) \rangle _{\beta=0} 
 =  \frac{  \int d U_\mu(s) \cos \theta_\mu(s)}
 {  \int d U_\mu(s)}  
 = 
\frac{\int_{0}^{\frac{\pi}{2}} d \theta_\mu 
\sin \theta_\mu  \cos^2 \theta_\mu }
 { \int_{0}^{\frac{\pi}{2}} d \theta_\mu  
\sin \theta_\mu \cos \theta_\mu  } =
\frac23.
\end{eqnarray}
In the MA gauge, we find $\langle R_{\rm Abel}\rangle_{\rm MA}
=\langle \frac12 {\rm Re}$ ${\rm tr}(U_{\mu}(s) 
u_{\mu}^\dagger(s)) \rangle \simeq 1$, 
and the SU(2) link variable is U(1)$_3$-like as 
$U_\mu(s) \simeq u_\mu(s)$ in the relevant gauge configuration. 
As a typical example, one obtains   
$\langle  R_{\rm Abel} \rangle_{\rm MA} \simeq$  0.926  on $16^4$ lattice
with $\beta = 2.4$.
Thus, in the MA gauge, the amplitude of the off-diagonal gluon 
$A_\mu^\pm(x)$ is strongly suppressed, 
which can be called as microscopic abelian dominance. 
On the other hand, 
the phase degrees of freedom $\tilde \chi_\mu(x)$ of 
$A_\mu^\pm(x)$ is not constrained by the MA gauge-fixing condition 
at all, and the constraint from the QCD action is also suppressed 
because of the strong reduction of $|A_\mu^\pm(x)|$ in the MA gauge. 
Therefore, in the MA gauge, the phase degrees of freedom $\tilde 
\chi_\mu(x)$ 
of the off-diagonal gluon $A_\mu^\pm(x)$ behaves as a random angle 
variable approximately, and this phase randomness leads to 
macroscopic abelian dominance on the confinement force \cite{ichiead}.

\subsection{Maximally Abelian Gauge in the Connection Formalism}

In the gauge theory, the covariant derivative is more fundamental 
than the gauge field, and therefore the MA gauge fixing in the 
continuum SU($N_c$) QCD using the SU($N_c$) covariant derivative 
operator ${\hat D_\mu} \equiv {\hat \partial_\mu} + ie A_\mu $, 
where $\hat \partial_\mu$ is the derivative operator 
satisfying $[\hat \partial_\mu, f(x)]= \partial_\mu f(x)$. 
In addition, both the derivative operator and the Lie algebra 
appearing in ${\hat D_\mu}$ are expressed by the infinitesimal 
transformation of the corresponding group elements, 
so that they are to be described by way of the commutation relation. 
Then, the MA gauge is defined so as to minimize 
\begin{eqnarray}
 R_{\vec H}[A_\mu(\cdot)] \equiv \int d^4 x \; {\rm tr}[{\hat D_\mu}, \vec 
H]^\dagger
[{\hat D_\mu}, \vec H]
= e^{2} \int d^{4}x \; {\rm tr}[A_\mu, \vec H]^{\dagger}[A_\mu, \vec H] 
\nonumber \\
= e^{2} \int d^4 x \; \sum_{\alpha, \beta}A_\mu^{\alpha*}A_\mu^{\beta}
\vec \alpha \cdot \vec \beta {\rm tr} (E_\alpha^{\dagger} E_\beta)
= \frac{e^{2}}{2} \int d^{4}x \sum_{\alpha = 1}^{N_c(N_{c}-1)}
|A_{\mu}^{\alpha}|^{2}
\label{eq:rch}
\end{eqnarray}
by the gauge transformation in the Euclidean QCD. 
Here, we have used the Cartan decomposition,
$\displaystyle
A_{\mu} \equiv A_\mu^{a} T^{a} = \vec A_{\mu}  \cdot \vec H +
\sum_{\alpha = 1}^{N_c(N_{c}-1)} A_{\mu} ^{\alpha} E^{\alpha}$, where  
$\vec H \equiv  (T_3, T_8, \cdots, T_{N_c^2-1})$ is the Cartan 
subalgebra, and $E^{\alpha}(\alpha=1,2,\cdots,N_{c}^{2}-N_c)$ 
denotes the raising or lowering operator.
Since $ R_{\vec H}[A_{\mu}]$ expresses the total amount of the off-diagonal
gluon component,
SU($N_{c}$) gauge connection $\hat D _{\mu}= \hat \partial_{\mu} + ie 
A_{\mu}^{a} T^{a}$ is mostly close to U(1)${}^{N_{c}-1}$ gauge connection 
$\hat D _{\mu}^{\vec H}= \hat \partial_{\mu} + ie \vec A_{\mu} \cdot 
\vec H$ in the MA gauge. 
In our definition  (\ref{eq:rch})  using $\hat D _{\mu}$,
the gauge transformation 
property of $ R_{\vec H}[A_{\mu}]$ becomes quite transparent, 
because the SU($N_c$) covariant derivative ${\hat D_\mu}$
obeys the simple adjoint gauge transformation, 
${\hat D_\mu} \rightarrow 
\Omega {\hat D_\mu} \Omega^{\dagger}$,
with the SU($N_c$) gauge function $\Omega \in$ SU($N_c$).
By the SU($N_c$) gauge transformation, $ R_{\vec H}$ is transformed as
 \begin{eqnarray}
 R_{\vec H} \rightarrow R_{\vec H}^{\Omega} =
 \int d^{4} x \; {\rm tr}
 \left( [\Omega \hat D_{\mu} \Omega^{\dagger}, \vec H]^{\dagger}
[\Omega \hat D_{\mu} \Omega^{\dagger}, \vec H] \right) \nonumber \\
=     \int d^{4} x \; {\rm tr} 
 \left( [\hat D_{\mu}, \Omega^{\dagger}\vec H \Omega]^{\dagger}
[\hat D_{\mu}, \Omega^{\dagger}\vec H \Omega]\right),
\end{eqnarray} 
and hence
the residual symmetry corresponding to the invariance of $ R_{\vec H}$  
is easily found to be U(1)$^{N_{c}-1}_{\rm local} \times P^{N_{c}}_{\rm 
global} \subset $SU($N_{c})_{\rm local}$, where  $P^{N_{c}}_{\rm 
global}$ denotes the global Weyl group relating to the permutation of 
the $N_{c}$ bases in the fundamental representation,
and $N_{c}!$ elements includes.
In fact, one finds $\omega^{\dagger} \vec H \omega = \vec H$ for 
$\omega = e^{-i \vec \varphi(x) \cdot \vec H} \in$   
U(1)$^{N_{c}-1}_{\rm local}$, and the global Weyl transformation by 
$W \in$ $P^{N_{c}}_{\rm global}$ only exchanges the permutation of the 
nontrivial root $\vec \alpha_{j}$ and never changes $ R_{\vec H}$.
In the MA gauge, by definition, arbitrary gauge transformation by 
${}^\forall V \in 
$ SU($N_{c}$) is to increase 
$ R_{\vec H}$ as $ R_{\vec H}^{V} \ge  R_{\vec H}$.
Considering arbitrary infinitesimal gauge transformation $V = 
e^{i \varepsilon} \simeq 1 + i \varepsilon $ with ${}^\forall \varepsilon 
\in$$su$($N_{c}$), one finds $V^{\dagger} \vec H V \simeq \vec 
H + i[\vec H, \varepsilon]$ and
\begin{eqnarray}
R_{\vec H}^{V} \simeq   R_{\vec H} + 2i\int d^{4}x 
{\rm tr} \left( [\hat D_{\mu}, [\vec H, \varepsilon]]^{\dagger} 
[\hat D_{\mu}, \vec H] \right)
\nonumber \\
=  R_{\vec H} + 2i\int d^{4}x 
{\rm tr}\left( \varepsilon
[\vec H, [\hat D_{\mu}^{\dagger},[\hat D_{\mu},\vec H] ]] \right).
\end{eqnarray}
In the MA gauge, the extremum condition of $ R_{\vec H}^{V}$ on
${}^\forall \varepsilon \in $$su$($N_{c}$) provides 
\begin{eqnarray}
[\vec H, [\hat D_{\mu}^{\dagger}, [\hat D_{\mu},\vec H] ]] = 0,
\end{eqnarray}
which leads to $\sum_{\mu}(i \partial_{\mu} \pm eA^{3}_{\mu}) 
A_{\mu}^{\pm} = 0$ for the $N_{c}$=2 case.
Thus, the variable to be diagonalized in the MA gauge is 
easily derived as 
\begin{eqnarray}
\vec \Phi[A_{\mu}] = [\hat D_{\mu}^{\dagger}, 
[\hat D_\mu, \vec H]]  \in su(N_{c})
\label{eq:Pdiag}
\end{eqnarray}
in the continuum theory.
Here, $\vec \Phi[A_{\mu}]$ is hermite as $\vec \Phi^\dagger [A_{\mu}]
=\vec \Phi[A_{\mu}]$ because of $\hat D_{\mu}^\dagger = -\hat 
D_{\mu}$,
and hence the diagonal elements of $\vec \Phi[A_{\mu}]$ should be real.

Thus, $\vec \Phi[A_{\mu}]$ can be regarded as a sort of the 
`gluonic Higgs field' relating to the MA gauge  fixing, however, 
$\vec \Phi(A_{\mu})$ does not obey the adjoint gauge transformation, so 
that correspondence between monopole and $\vec \Phi[A_{\mu}]$ is still 
unclear.
The deviation of the `gluonic Higgs field' $\phi[A_{\mu}]$ obeying the 
adjoint transformation will be discussed in section 5. 

In the commutator form, the diagonal part of the variable 
$\hat O[A_\mu(x)]$ is expressed as 
\begin{eqnarray} 
\hat O^{\vec H} = \hat O - [\vec H, [\vec H, \hat O]]. 
\label{eq:APdif}
\end{eqnarray}
For the covariant derivative operator, one finds 
\begin{eqnarray} 
\hat D_\mu^{\vec H} = \hat D_\mu - [\vec H, [\vec H, \hat D_\mu]] 
=\hat \partial_\mu+ie \vec A_\mu(x) \cdot \vec H 
\end{eqnarray}
with $A_\mu(x)=\vec A_\mu(x) \cdot \vec H+A_\mu^\alpha(x)E^\alpha$. 
Then, the abelian projection, 
$\hat D_\mu \rightarrow \hat D_\mu^{\vec H}$, is expressed 
by the simple replacement as 
$A_\mu(x) \in su(N_c) \rightarrow {\cal A}_\mu(x) \equiv
 \vec A_\mu(x) \cdot \vec H$ $\in$ $u(1)^{N_{c}-1}$. 

\subsection{Generalization of the Maximally Abelian Gauge}

In the MA gauge, $R_{\vec H}[A_\mu(\cdot)]$ in Eq.(\ref{eq:rch}) is 
forced to be reduced by the MA gauge transformation 
$\Omega_{\rm MA}(x) \in G/H$\cite{ichiead},
and therefore the gluon field $A_\mu(x)$ 
is maximally arranged in the diagonal direction 
$\vec H$ in the internal SU$(N_c)$ color space. 
In the definition of the MA gauge, $\vec H$ is the 
specific color-direction, 
since $\vec H$ explicitly appears in the MA gauge-fixing 
condition with $R_{\vec H}[A_\mu(\cdot)]$. 
On this point of view, the MA gauge can be called as the 
`maximally diagonal gauge'. 
However, for the extraction of the abelian gauge theory from the 
nonabelian theory, 
we need not take the specific direction as $\vec H$ in the internal 
color-space, although the system becomes transparent 
when the specific color-direction as $\vec H$ is introduced 
on the maximal arrangement of the gluon field $A_\mu(x)$. 

In this subsection, we consider the generalization of the 
framework of the MA gauge and the abelian projection, 
without explicit use of the specific direction $\vec H$
in the internal color-space 
on the gauge fixing.
(Such an attempt is similar to the generalization of 
the formalism in the center-of-mass frame to 
that in the general moving frame.) 
Instead of the special color-direction $\vec H$,
we introduce the 
`Cartan frame field' 
$\vec \phi(x) \equiv (\phi_1(x), \phi_2(x), \cdots, \phi_{N_c-1}(x))$,
where $\phi_i(x) \equiv \phi^a_i(x) T^a$ $(\phi_i^a(x) \in {\bf R})$
commutes each other as $[\phi_i(x), \phi_j(x)]=0$, and satisfy the
orthonormality condition
2tr$(\phi_i(x) \phi_j(x))=\sum_{a=1}^{N_c-1} \phi^a_i(x) \phi^a_j(x) 
= \delta_{ij}$. 
At each point $x_\mu$, $\vec \phi(x)$ forms the Cartan sub-algebra, and 
can be expressed as
\begin{eqnarray}
\vec \phi(x) = \Omega_C^\dagger(x) \vec H \Omega_C(x)
\end{eqnarray} 
using $\Omega_C(x)$ $\in$ $G/H$.
For the fixed Cartan frame field $\vec \phi(x)$, 
we define the generalized maximally abelian (GMA) gauge 
so as to minimize the functional 
\begin{eqnarray}
R_{\phi}[A_\mu(\cdot)] \equiv \int d^4 x 
{\rm tr}[{\hat D_\mu}, \vec \phi(x)]^\dagger 
[{\hat D_\mu}, \vec \phi(x)]
\end{eqnarray}
by the SU$(N_c)$ gauge transformation.
Here, the Cartan frame field $\vec \phi(x)$ is defined at each $x_\mu$
independent of the gluon field like $\vec H$, and never changes under the 
SU($N_c$) gauge transformation.
For the special case of $\vec \phi(x)=\vec H$, 
the GMA gauge returns to the usual MA gauge. 
In the GMA gauge, the SU($N_c$) covariant derivative $\hat D_\mu$ 
is maximally arranged to be `parallel' 
to the $\vec \phi(x)$-direction in the internal color-space 
using the SU($N_c$) gauge transformation.

In the GMA gauge, the gauge symmetry 
is reduced from SU($N_c$) 
into ${\rm U}(1)_\phi^{N_c-1}$, and
the generalized AP-QCD leads to the monopole 
in the similar manner to the MA gauge. 
In the GMA gauge,  the remaining U(1)$^{N_c-1}_{\phi}$ gauge symmetry 
corresponds to the invariance of $R_{\phi}[A_\mu(\cdot)]$ 
under the U(1)$^{N_c-1}_{\phi}$ gauge transformation by 
\begin{eqnarray}
\omega_\phi(x) \equiv e^{i \vec \phi(x) \cdot 
\vec \chi(x)} \in {\rm U}(1)_\phi^{N_c-1},
\qquad  \vec \chi(x) \in {\bf R}^{N_c-1}.
\end{eqnarray}
In fact, using
$\omega_{\phi}^\dagger(x) \vec \phi(x) \omega_{\phi}(x) 
= \vec \phi(x)$, U(1)$^{N_c-1}_\phi$ invariance of $R_\phi[A_\mu(\cdot)]$
is easily confirmed as
\begin{eqnarray}
(R_\phi[A_\mu])^\omega
&  = & \int d^4x {\rm tr}
[\omega(x)_{\phi}\hat D_\mu \omega^{\dagger}_{\phi}(x), \vec \phi(x)]^\dagger
[\omega(x)_{\phi} \hat D_\mu \omega^{\dagger}_{\phi}(x), \vec \phi(x)]   
\\
&  = & \int d^4x {\rm tr}
[\hat D_\mu , \omega_{\phi}^\dagger (x) \vec \phi(x) \omega_{\phi}(x)]^\dagger
[\hat D_\mu , \omega_{\phi}^\dagger (x) \vec \phi(x) \omega_{\phi}(x)] 
  =  R_\phi[A_\mu]. \nonumber 
\end{eqnarray}
There also remains the global Weyl symmetry ${\bf P}_{N_{c}}$
similarly in the usual MA gauge,
although the gauge function takes a complicated from.

Here, we consider the generalized abelian projection 
to $\vec \phi(x)$-direction. 
Similar to the `diagonal part' in Eq.(\ref{eq:APdif}), 
we define the `$\vec \phi(x)$-projection' of the operator $\hat O(x)$ as 
\begin{eqnarray} 
\hat O^{\phi}(x) = \hat O(x) - [\vec \phi(x), [\vec \phi(x), \hat 
O(x)]],
\end{eqnarray}
using the commutation relation.
For the SU($N_c$) covariant derivative operator $\hat D_\mu \equiv \hat 
\partial_\mu + ie A_\mu$, its $\vec \phi(x)$-projection is defined as 
\begin{eqnarray} 
\hat D_\mu^\phi \equiv \hat D_\mu -[\vec \phi(x), [\vec \phi(x), 
\hat D_\mu]]= \hat \partial_\mu + ie {\cal A}_\mu^\phi(x) + 
[\vec \phi(x), \partial_\mu \vec \phi(x)]
\end{eqnarray}
with ${\cal A}_{\mu}^\phi(x) \equiv \vec A_{\mu}^\phi(x) \cdot \vec 
\phi = 2 {\rm tr}(\vec \phi(x) A_{\mu}(x)) 
\cdot \vec \phi(x)$.
Here, the nontrivial term $[\vec \phi(s), \partial_\mu \vec \phi(x)]$ 
appears in $\hat D_\mu^\phi$ owing to the $x$-dependence of the Cartan-frame
field $\vec \phi(x).$
The U(1)$^{N_c-1}_\phi$ gauge field is defined as the difference 
between $\hat D^\phi_\mu $ and $\hat \partial_\mu$,
\begin{eqnarray} 
\tilde {\cal A}^\phi_\mu(x) \equiv \frac{1}{ie}(\hat D^\phi_\mu - \hat 
\partial_\mu) = {\cal A}_\mu^\phi(x) + \frac{1}{ie}[\vec \phi(x), \partial_\mu 
\vec \phi(x)] \hspace{.5cm} \in \hspace{.5cm} su(N_c).
\end{eqnarray}
Here, $\tilde {\cal A}_\mu^\phi(x)$ includes both the $\vec 
\phi(x)$-component 
${\cal A}_\mu^\phi(x) = 2 {\rm tr}( A_\mu(x) \vec \phi(x)) \cdot \vec \phi(x)$ 
and the non-$\vec \phi(x)$-component  $\frac{1}{ie}[\vec \phi(x), \partial_\mu 
\vec \phi(x)]$, because $[\vec \phi(x), \partial_\mu \vec \phi(x)]$ 
does not include $\vec \phi(x)$-component as
${\rm tr} \left( \phi_i(x) [\vec \phi(x), \partial_\mu \vec \phi(x)] 
\right)=0$. 
Here, the $\vec A_\mu^\phi(x)$ is the image of 
$\tilde A_\mu^\phi(x)$ mapped into the U(1)$_\phi^{N_c-1}$-manifold. 
The the generalized abelian projection for the variable $O[A_\mu(x)]$ 
is defined via the two successive mapping, 
$O[A_\mu(x)] \rightarrow O[\tilde {\cal A}_\mu^\phi(x)] 
 \rightarrow \vec O_{AP} \equiv 
 2{\rm tr}(\vec \phi(x) O[\tilde {\cal A}_\mu^\phi(x)])$, 
after the GMA gauge fixing.

Under the U$(1)_\phi^{N_c-1}$ abelian gauge transformation by 
$\omega_\phi(x)=e^{i \vec \phi(x) \cdot \vec \chi(x)} \in 
{\rm U}(1)_\phi^{N_c-1}$, 
$\tilde {\cal A}_\mu^\phi(x)$ or $\vec A_\mu^\phi(x)$ 
behaves as the U(1)$^{N_c-1}_\phi$ abelian gauge field,
\begin{eqnarray}
\tilde {\cal A}^\phi_\mu(x) \rightarrow (\tilde {\cal A}^\phi_\mu(x))^\omega
= \tilde {\cal A}_\mu^\phi(x) + \frac{1}{e} \partial_\mu \vec \chi_\mu(x) \cdot
\vec \phi(x).
\end{eqnarray}
The U(1)$_\phi^{N_c-1}$-abelian field strength is defined as 
$\tilde {\cal F}_{\mu\nu}^\phi(x) \equiv 
\frac{1}{ie}\{[\hat D_\mu^\phi, \hat D_\nu^\phi]
-[\hat \partial_\mu, \hat \partial_\nu]\},$ 
which generally includes the non-$\vec \phi(x)$ component as well as 
$\tilde {\cal A}_\mu^\phi(x)$. 
In the U(1)$_\phi^{N_c-1}$-manifold, 
the $\vec \phi(x)$-component 
$
\vec F_{\mu\nu}^\phi(x) \equiv  2 {\rm tr}
\left( \tilde {\cal F}_{\mu\nu}^\phi(x) \vec \phi(x) \right) 
$
is observed as the mapped image of $\tilde {\cal F}_{\mu\nu}^\phi(x)$.

Next, we investigate the properties of the GMA gauge function $\Omega_{\rm 
GMA}(x)$,
which brings the GMA gauge.
Here, $\Omega_{\rm GMA}(x)$ is a complicated function of $A_\mu(x)$ and is 
expressed by an element of the coset space 
$G/H= {\rm SU}(N_c)/ \{ {\rm U}(1)^{N_c-1}_\phi \times
{\rm Weyl} \} $ as the representative element 
because of the residual gauge symmetry.
For instance, we impose here
\begin{eqnarray}
{\rm tr}(\Omega_{\rm GMA}(x) \vec \phi(x)) = \vec 0
\end{eqnarray}
for the selection of $\Omega_{\rm GMA}$ $\in$ $G/H$. 
Similarly to the MA gauge function\cite{ichiead}, 
$\Omega_{\rm GMA}[A_\mu]$ obeys
the nonlinear transformation as
\begin{eqnarray}
\Omega_{\rm GMA}(x) \in G/H \rightarrow (\Omega_{\rm GMA}(x))^V =
d^V(x)\Omega_{\rm GMA}(x)V^\dagger(x) \,\, \in \,\, G/H 
\end{eqnarray}
by the SU($N_c$) gauge transformation with $V(x)$ $\in$ $G$.
Here, $d^V(x) \in  H
\equiv$ U(1)$^{N_{c}-1}_\phi$ $\times$ Weyl 
appears to keep $(\Omega_{\rm GMA})^V$ belonging to $G/H$.
Therefore, the gluon field 
$A^{\rm GMA}_\mu = \Omega_{\rm GMA}(A_\mu+\frac{1}{ie}
\partial_\mu)
\Omega^\dagger_{\rm GMA} $ $\in$ $g$ in the GMA gauge is transformed as
\begin{eqnarray}
A_\mu^{\rm GMA} \rightarrow (A_\mu^{\rm GMA})^V  & = &
\Omega^V_{\rm GMA}(x)
(A_\mu^V+\frac{1}{ie}\partial_\mu) \Omega_{\rm GMA}^{V\dagger}(x) 
\nonumber \\
& = & d^V(x) (A_\mu^{\rm GMA}+\frac{1}{ie}\partial_\mu) d^{V\dagger}(x) = 
(A_\mu^{\rm GMA})^{d^V}
\label{eq:subabelian}
\end{eqnarray}
by the SU($N_c$) gauge transformation.
As a remarkable feature, the SU$(N_c)$ gauge transformation by $V(x)$ 
$\in$ $G$ is mapped as 
the abelian  sub-gauge transformation by $d^V(x)$ $\in$ $H$  in the GMA gauge:
$(A_\mu^{\rm GMA})^V = (A_\mu^{\rm GMA})^{d^V}$. 
In particular, for the residual gauge transformation by $\omega(x) = 
e^{i \vec \phi(x) \cdot \vec \chi(x)}$ $\in$
$H$, we find $d^\omega(x) = \omega(x)$ to keep the representative-element 
condition
tr$(\Omega^\omega_{\rm GMA}(x) 
\vec \phi(x)) = \vec 0$ imposed above, 
and then $A_\mu^{\rm GMA}$ obeys the ordinary $H$-gauge 
transformation
\begin{eqnarray}
A_\mu^{\rm GMA}(x) \rightarrow (A_\mu^{\rm GMA}(x))^\omega = 
\omega(x) (A_\mu^{\rm GMA}+\frac{1}{ie}\partial_\mu) 
\omega^{\dagger}(x).
\end{eqnarray}
For the arbitrary variable $\hat O[A_\mu^{\rm GMA}]\equiv \hat 
O[A_\mu^{\Omega_{\rm GMA}}]$ defined in the GMA gauge, 
we find $\hat O[A_\mu^{\rm GMA}]^V =
\hat O[A_\mu^{\rm GMA}]^{d^V} $ with $d^V$ $\in$ $H$ from 
Eq.(\ref{eq:subabelian}),
and hence we get an useful criterion on the SU($N_c$) gauge invariance:
if $\hat O[A_\mu]$ is $H$-invariant as 
$\hat O[A_\mu^{\rm GMA}]^\omega = \hat O[A_\mu]$ for ${}^\forall \omega$ 
$\in$ $H$, $\hat O[A_\mu^{\rm GMA}]$ is also $G$-invariant, because of 
$\hat O[A_\mu^{\rm GMA}]^V = 
\hat O[A_\mu^{\rm GMA}]^{d^V} = \hat O[A_\mu^{GMA}]$
for ${}^\forall V$ $\in$ $G$.
All of the above arguments are also applicable to the usual MA gauge by 
setting $\vec \phi(x)= \vec H$.

For the regular field $\vec \phi(x)$ without any discontinuity, 
the GMA gauge function $\Omega_{\rm GMA}$ $\in$ SU($N_c$)/U(1)$^{N_c-1}_\phi$
becomes singular like the MA gauge, which was discussed in section 2.3.
Then, a nontrivial singular term appears in the field strength as 
\begin{eqnarray}
G_{\mu \nu }^{\rm GMA}= \partial_\mu A_\nu ^{\rm GMA} -\partial_\nu A_\mu ^{\rm GMA}
 + ie[A_\mu ^{\rm GMA} , A_\nu ^{\rm GMA} ] 
+ \frac{i}{e} \Omega _{\rm GMA} [\partial_\mu , \partial_\nu ]
 \Omega _{\rm GMA}^\dagger.
\end{eqnarray}
Similar to the MA gauge, the singularity on $\Omega _{\rm GMA}$ induces 
breaking of the U(1)$_\phi^{N_c-1}$ abelian Bianchi 
identity and the monopole current 
in the U(1)$_\phi^{N_c-1}$ abelian sector. 

The correspondence between $\Omega _{\rm GMA}$ and 
$\Omega _{\rm MA}$ is 
straightforward.  
Using  
$\Omega _C(x) \in {\rm SU}(N_{c})$ satisfying 
$\vec \phi(x)=\Omega _C^\dagger(x) \vec H \Omega_C(x)$,
$\Omega _{\rm GMA}$ is expressed as 
\begin{eqnarray}
\Omega^{\rm GMA}(x)=\Omega _C^\dagger(x)\Omega ^{\rm MA}(x).
\end{eqnarray}
Then, for regular $\vec \phi (x)$, $\Omega _C(x)$ becomes
regular, and the singularity of $\Omega _{\rm MA}$ 
is directly mapped to that of $\Omega _{\rm GMA}$. 
However, if singular $\vec \phi(x)$ is used, 
the singularity of $\Omega _{\rm MA}$ can be mapped in $\vec \phi(x)$
 or $\Omega_C(x)$
instead of $\Omega _{\rm GMA}$.
In this case, the gluon field $A_\mu^{\rm GMA}$ is kept to be regular, and 
the Cartan frame field
 $\vec \phi(x)$ includes the multi-valuedness or the 
singularity, which leads to the monopole. 
Such a situation will be discussed in section 5 considering the analogy with 
the nonabelian Higgs theory.

\section{Large Field Fluctuation around Monopoles}

In this section, we study the QCD-monopole appearing 
in the abelian gauge in terms of the gluon field 
fluctuation\cite{ichiem}.
For simplicity, we take $N_c=2$.
In the static frame of the QCD-monopole 
with the magnetic charge $g$, 
a spherical `magnetic field' is created around the monopole 
in the abelian sector of QCD as 
\begin{eqnarray}
{\bf H}(r) =  \frac{g}{4\pi r^3} {\bf r}
\end{eqnarray}
with ${\bf H}_i\equiv  \varepsilon _{ijk} \partial_j A^3_k$. 
Thus, the QCD-monopole inevitably accompanies 
a large fluctuation of the abelian gluon component $A^3_\mu$ around it. 
As was discussed in section 2, in the abelian gauge, 
the formal action of the abelian projected QCD 
or the abelian part of the QCD action is given by 
$S^{\rm Abel}\equiv -\frac14 \int d^4x
\{ (\partial_\mu A_\nu^3-\partial_\nu A_\mu^3)^2  
-F_{\mu\nu}^{\rm sing}\}$, where 
$-F_{\mu\nu}^{\rm sing}$ appears and 
eliminates the Dirac-string contribution. 
In the abelian part, 
the field energy created around the monopole 
is estimated as the ordinary electro-magnetic energy, 
\begin{eqnarray}
{\cal E}(a) = \int_a^\infty d^3 x \frac12 {\bf H}(r)^2 = 
\frac{g^2}{ 8 \pi a},
\label{monopolemass1}
\end{eqnarray}
where $a$ is an ultraviolet cutoff like a lattice mesh.
As the `mesh' $a$ goes to 0, 
the monopole inevitably accompanies an infinitely large 
energy-fluctuation in the abelian part 
and makes $S^{\rm Abel}$ divergent. 

Since there seems no plausible reason to eliminate 
such a divergence via renormalization, 
the monopole seems difficult to appear in the abelian 
gauge theory controlled by $S^{\rm Abel}$.
This is the reason why QED does not have the point-like Dirac 
monopole.
Then, why can the QCD-monopole appear in the abelian projected QCD ? 
To answer it, let us consider the division of the total QCD action 
$S^{\rm QCD}$ into the abelian part $S^{\rm Abel}$ and 
the remaining part $S^{\rm off}\equiv S^{\rm QCD}-S^{\rm Abel}$, which 
is contribution from the off-diagonal gluon component. 
While $S^{\rm QCD}$ and $S^{\rm Abel}$ are positive definite in the 
Euclidean metric, 
$S^{\rm off}$ is not positive definite and can take a negative value.
Then, around the QCD-monopole, 
the abelian action $S^{\rm Abel}$ should be 
partially canceled by the remaining contribution $S^{\rm off}$ from 
the off-diagonal gluon component, so as to keep 
the total QCD action $S^{\rm QCD}$ 
%
%
finite even for $a\rightarrow 
0$. 
Similar cancellation between the gauge field and the Higgs field 
fluctuation is also found around the GUT monopole. 
Thus, we expect large off-diagonal gluon components 
around the QCD-monopole for its existence as well as a 
large field fluctuation in the abelian part.
Based on this analytical consideration, we study the field fluctuation and 
monopoles in the MA gauge using the lattice QCD.

To begin with, 
we study the gluon field configuration around the monopole 
in the MA gauge in terms of 
%
%
the abelian angle variable $\theta_\mu^3(s)$ and abelian projection rate
$R_{\rm Abel} \equiv \cos \theta_\mu(s)$, which measures 
the off-diagonal gluon remaining in the MA gauge \cite{ichiead}.
For the argument of $\theta^3_\mu(s)$, 
%
%
the U(1)$_3$ gauge degrees of freedom should be also fixed after the 
MA gauge fixing, because $\theta_\mu^3(s)$ is U(1)$_3$ gauge dependent. 
%
%
Here, we adopt the U(1)$_3$ Landau gauge \cite{mandula,amemiya} defined by maximizing
\begin{eqnarray}
R[U_\mu] \equiv \sum_{s,\mu}  {\rm tr} u_\mu(s) 
=2 \sum_{s,\mu} \cos \theta_\mu^3(s) 
\end{eqnarray}
by the residual U(1)$_3$ gauge transformation.
In the U(1)$_3$ Landau gauge, there remains no local symmetry, and 
the lattice variable mostly approaches to the continuum field 
under the constraint of the MA gauge fixing.

Now, let us consider the correlation between the field variables 
and the monopole in the lattice QCD. 
For this argument, one has to recall the property of the monopole current 
shown in section 2.4. 
In particular, one should note that $k_\mu(s)$ is defined on the dual 
link and only affects the perpendicular components to the $\hat \mu$-direction
 for the electric variable as $F_{\alpha\beta}$ because of
$k_\mu \equiv \partial_\alpha^* F_{\alpha\mu} = -\frac12
\varepsilon_{\mu\alpha\beta\gamma} \partial_\alpha F_{\beta\gamma}$.

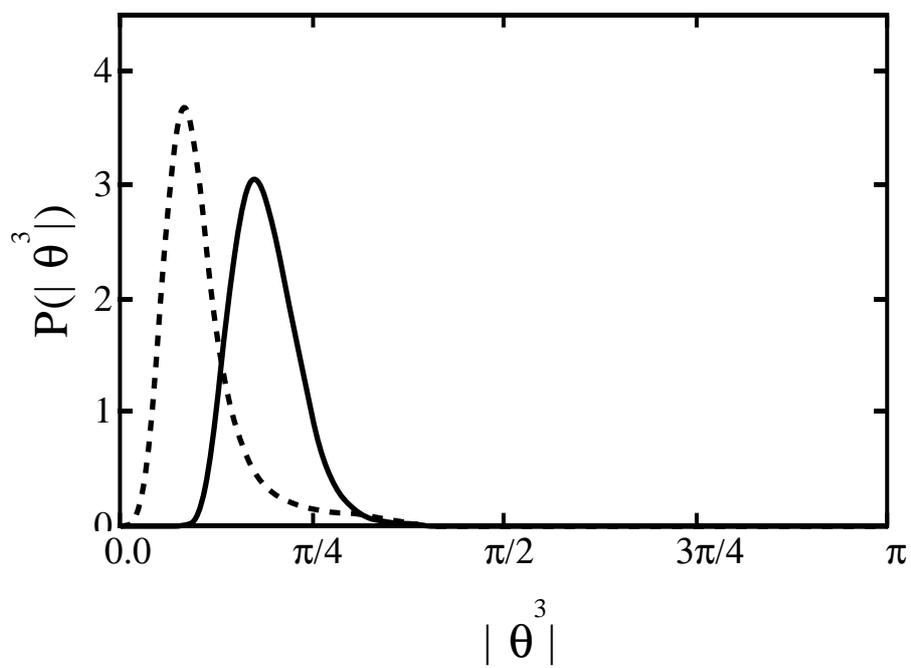
\begin{figure}[bt]
\caption{
The solid curve denotes the 
probability distribution $P( |\bar \theta^3|)$ of the 
averaged abelian angle variable $|\bar \theta^3(s, \hat \mu)|$ 
around the monopole current
in the MA gauge with the U(1)$_3$ Landau gauge fixing.
Here, $ |\bar \theta^3(s,\hat \mu)|$ is the average of 
$|\theta_\alpha^3(s)|$ over the 
neighboring links around the dual link,
and data are obtained from the SU(2) lattice QCD with $\beta=2.4$ 
and $16^4$.
For comparison, the total distribution $P$ on the whole lattice is 
also added by the dashed curve.
Around the monopole, $ |\bar \theta^3|$ corresponding to the abelian 
gluon component takes a large value.
}
\label{gfig8}
\vspace{0cm}
\end{figure}

Taking account of these properties, 
we study the local correlation between the field variables 
and the monopole current $k_\mu(s)$ 
in the MA gauge with the U(1)$_3$ Landau gauge 
using the lattice-QCD Monte-Carlo simulation. 
We first measure the average of the abelian angle 
variable $\theta^3_\mu(s)$ over the neighboring links 
around the dual link (See Fig.\ref{monopole}),
\begin{eqnarray}
 |\bar \theta^3(s,\hat \mu)| \equiv \frac{1}{12} \sum_{\alpha\beta\gamma} 
\sum_{m,n=0}^{1} 
\frac12 | \varepsilon_{\mu\alpha\beta\gamma}| \cdot 
|\theta^3_\alpha (s+m \hat \beta + n \hat \gamma) |,
\end{eqnarray} 
which only consists of the perpendicular components considering 
the above monopole property.
%
%
Here, the index $\hat \mu$ denotes the direction of the dual link, 
and $|\bar \theta^3(s,\hat \mu)|$ corresponds to the average over the 12 
sides of the 3-dimensional cube perpendicular to the 
$\hat \mu$-direction.
We show in Fig.\ref{gfig8} the probability distribution $P(|\bar \theta^3|)$ of 
$|\bar \theta^3(s,\hat \mu)|$ in the MA gauge with the U(1)$_{3}$ Landau gauge at 
$\beta = 2.4$.
The solid curve denotes $P(|\bar \theta^3|)$ 
around the monopole current,
while the dashed curve denotes the total distribution 
on the whole lattice. The abelian angle variable $|\theta^3_\mu(s)|$ 
takes a large value around the monopole. 
In other words, the monopole provides the large fluctuation of 
the  abelian gauge field, which would enhance the randomness 
of the abelian link variable.

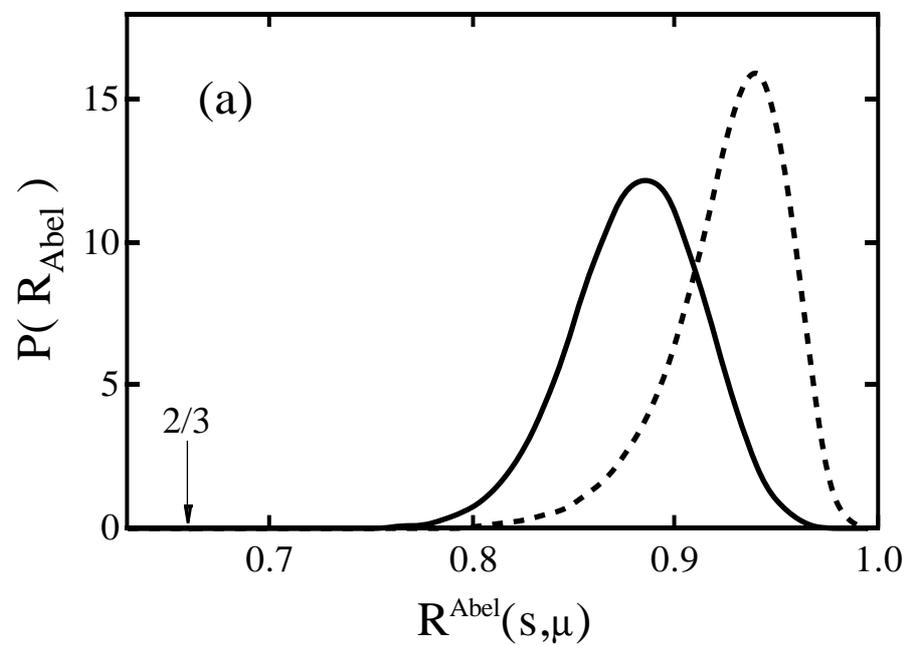
\begin{figure}[bt]
\caption{
(a) The solid curve denotes the 
probability distribution $P(\bar R_{\rm Abel})$
of the averaged abelian projection rate $\bar R_{\rm Abel}(s,\hat \mu)$
 around the monopole current
in the MA gauge in the SU(2) lattice QCD with  $\beta =2.4$ on $16^4$.
For comparison, the total distribution $P$ on the whole lattice is 
also added by the dashed curve.
(b) The solid curve denotes abelian projection rate 
$\langle \bar R_{\rm Abel} \rangle $
around the monopole current in the MA gauge as the function of $\beta$.
The dashed curve denotes 
$\langle \bar R_{\rm Abel} \rangle $ on the whole lattice. 
}
\label{gfig9}
\vspace{0cm}
\end{figure}

Similar to $|\bar \theta^3(s,\hat \mu)|$, 
we measure the average $\bar R_{\rm Abel}$ of the 
abelian projection rate $R_{\rm Abel}(s,\hat \mu)\equiv \cos\theta_\mu(s)$
over the neighboring links 
around the dual link, 
\begin{eqnarray}
\bar R_{\rm Abel} (s,\hat \mu) \equiv \frac{1}{12} \sum_{\alpha\beta\gamma} 
\sum_{m,n=0}^{1} 
\frac12 | \varepsilon_{\mu\alpha\beta\gamma} | \cos \theta_\alpha 
(s+m \hat \beta + n \hat \gamma)
\end{eqnarray} 
in the MA gauge 
to investigate the correlation between off-diagonal gluons and monopoles.
As shown in Fig.\ref{gfig9}(a), 
$\bar R_{\rm Abel}$ around the monopole current becomes smaller
than the total average of $\bar R_{\rm Abel}$ 
and therefore 
{\it the magnitude of the off-diagonal gluon component becomes larger around 
the monopole}.
The $\beta$ dependence of the abelian projection rate $\langle R_{\rm Abel} 
\rangle$
is shown in  Fig.\ref{gfig9}(b).
Although $\langle R_{\rm Abel} \rangle$  on the whole lattice 
approaches to unity as 
$\beta \rightarrow \infty$,  
$\langle R_{\rm Abel} \rangle$ around the monopole is about 0.88
and is not changed even in the large $\beta$ region.
Thus, the monopole provides the large fluctuation both for 
the abelian 
field and for the off-diagonal gluon. 


We next study monopoles in terms of the plaquette action density.
We define the SU(2), abelian  and 
`off-diagonal'  plaquette action densities as
\begin{eqnarray}
S^{\rm SU(2)}_{\mu\nu} (s) & \equiv & 1 - \frac12 {\rm tr} 
\Box^{\rm SU(2)}_{\mu\nu}(s),  \\
S^{\rm Abel}_{\mu\nu} (s)   &  \equiv  &
1 - \frac12 {\rm tr} \Box^{\rm Abel}_{\mu\nu}(s), \\
S^{\rm off}_{\mu\nu} (s)  & \equiv  &
 S^{\rm SU(2)}_{\mu\nu}(s) - S^{\rm Abel}_{\mu\nu}(s),
\label{eq:abaction}
\end{eqnarray}
where $\Box^{\rm SU(2)}_{\mu\nu}(s)$ and 
$\Box^{\rm Abel}_{\mu\nu}(s)$ denote the SU(2) and the abelian plaquette 
variables, respectively; 
\begin{eqnarray}
 \Box^{\rm SU(2)}_{\mu\nu}(s) & \equiv &
U_\mu(s) U_\nu (s+ \hat \mu) U^{\dagger}_\mu(s+\hat \nu)
U^{\dagger}_\nu(s), \\
\Box^{\rm Abel}_{\mu\nu}(s) & \equiv &
u_\mu(s) u_\nu(s + \hat \mu) u^{\dagger}_\mu(s + \hat \nu)
u^{\dagger}_\nu(s).
\end{eqnarray}
Here, all of $S_{\mu\nu}$ are defined as symmetric tensors, 
$S_{\mu\nu}=S_{\nu\mu}$, 
instead of the Lorentz scalar, considering the above property  
of the monopole current. 
In the continuum limit $a \rightarrow 0$, $S^{\rm SU(2)}_{\mu\nu} (s)$ and
$S^{\rm Abel}_{\mu\nu} (s)$ are related to the SU(2) and the abelian 
action densities as 
$S^{\rm SU(2)}_{\mu\nu} (s) \rightarrow \frac14 a^4 e^2  {\rm tr} 
G^2_{\mu\nu}$ and
$S^{\rm Abel}_{\mu\nu} (s) \rightarrow \frac14 a^4 e^2  {\rm tr} {\cal 
F}^2_{\mu\nu}$, 
and then we call $S_{\mu\nu}$ as the action density, 
in spite of the lack of the summation on the Lorentz indices. 
Here, $S^{\rm off}_{\mu\nu}$ corresponds to the contribution of the
 off-diagonal gluon. 
While $S^{\rm SU(2)}_{\mu\nu}$ and $S^{\rm Abel}_{\mu\nu}$
are positive-definite, 
$S^{\rm off}_{\mu\nu}$ is not positive-definite and can take a 
negative value.

In order to examine the correlation between the action densities and 
the monopole current defined on the dual link, we measure the average of 
the action density $ S(s)$ over the neighboring plaquettes 
around the dual link, 
\begin{eqnarray}
\bar S(s,\hat \mu) \equiv
\frac16 \sum_{\alpha\beta\gamma} \sum_{m=0}^1 \frac12 
| \varepsilon_{\mu\alpha\beta\gamma} |
S_{\alpha\beta}(s + m \hat \gamma).
\end{eqnarray}
%
%
Here, $\hat \mu$ appearing in $\bar S(s,\hat \mu)$ denotes 
the direction of the dual link, and $\bar S(s,\hat \mu)$ corresponds to 
the average over 6 faces of the 3-dimensional cube perpendicular to 
the $\hat \mu$-direction. 

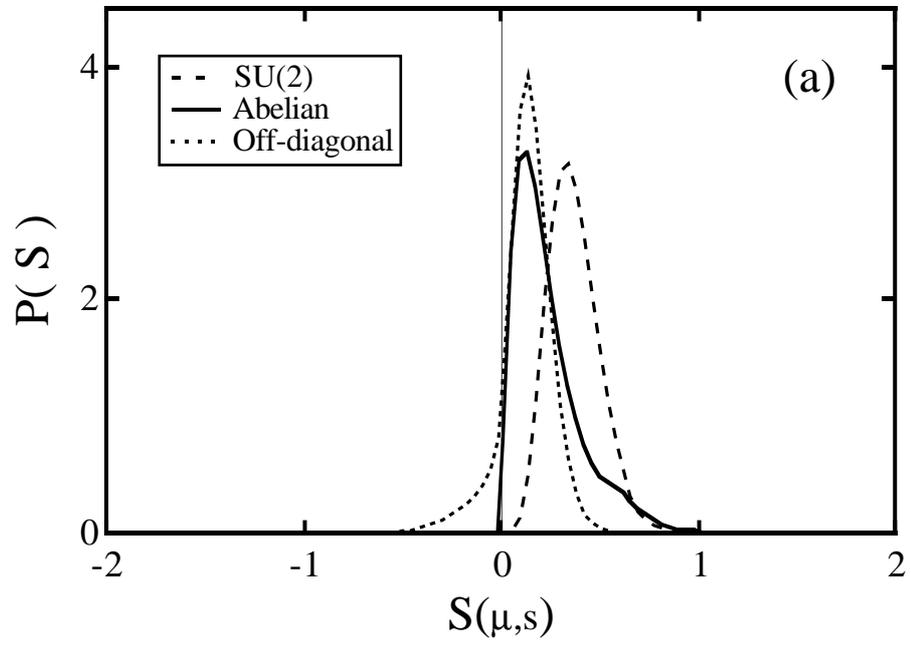
\begin{figure}[bt]
\caption{
(a) The probability distribution $P(\bar S)$ of 
density $\bar S(s,\hat \mu)$ on the whole lattice
in the MA gauge at $\beta =2.4$ on $16^4$ lattice.
(b)The probability distribution $P_k(\bar S)$ 
of the action density $\bar S(s,\hat \mu)$ around the monopole 
current ${k_\mu}$.
The dotted and the solid curves denote
$P(\bar S^{\rm SU(2)})$ and  $P(\bar S^{\rm Abel})$, respectively.
The dashed curve denotes  
$P(\bar S^{\rm off})$ for the off-diagonal part 
$\bar S^{\rm off}$ of the action density. 
}
\label{gfig10}
\vspace{0cm}
\end{figure}

We show in Fig.\ref{gfig10}
the probability distribution $P(\bar S)$ of the action densities
$\bar S(s,\hat \mu)$ in the SU(2), 
the abelian and the off-diagonal parts. 
Before the argument around the monopole current, 
we show the action densities 
on the whole lattice in Fig.\ref{gfig10} (a).
On the whole lattice, most 
$\bar S^{\rm off}$ are positive, and both $\bar S^{\rm Abel}$ and 
$\bar S^{\rm off}$ tend to take smaller values than
 $\bar S^{\rm SU(2)} = \bar S^{\rm Abel} +  \bar S^{\rm off}$. 
In other words, $\bar S^{\rm Abel}$ and positive $\bar S^{\rm off}$
additionally contribute to $\bar S^{\rm SU(2)}$.

However,  such a tendency of the action densities is drastically changed 
around the monopole as shown in Fig.\ref{gfig10}(b). 
We find remarkable features of 
the action densities around the monopole as follows.
\begin{enumerate}
\item 
Around monopoles, most $\bar S^{\rm off}$ take negative values,
and $\bar S^{\rm Abel}$ is larger than 
$\bar S^{\rm SU(2)} = \bar S^{\rm Abel} +  \bar S^{\rm off}$. 
\item 
Due to the cancellation between  $\bar S^{\rm Abel}$ and $\bar S^{\rm 
off}$, $\bar S^{\rm SU(2)}$ does not take an extremely 
large value around the monopole. 
\end{enumerate}
Thus, the large abelian action density $S^{\rm Abel}$ around the monopole 
is strongly canceled by the off-diagonal contribution $S^{\rm off}$ to 
keep
the total QCD action $S^{\rm QCD}=S^{\rm Abel}+S^{\rm off}$ small.
Here, different from $S^{\rm SU(2)}$, $S^{\rm Abel}$ itself does not 
control the system directly, and hence there is no severe constraint from 
$S^{\rm Abel}$. However, large $S^{\rm Abel}$ is still not preferable, 
because the large-cancellation requirement between $S^{\rm Abel}$
and $S^{\rm off}$ leads to a strong constraint on the off-diagonal gluon 
and brings the strong reduction of the configuration number.

Around the monopole, the abelian action density $S^{\rm Abel}$ 
takes a large value, and this value can be estimated from a following 
simple calculation.
Without loss of generality, the monopole-current direction is locally 
set to be parallel to the temporal direction as 
$k^{lat}_0 (s)  \equiv  \frac{1}{2\pi} \partial_\alpha {}^*{\bar 
\theta_{\alpha 0}} (s) 
= \pm 1$. 
Here, $k_0^{lat}(s)$ is expressed as the sum of 
six plaquette variables $\bar \theta_{ij}$ ($i,j$=1,2,3) 
around the monopole, because of 
$k_0^{lat}(s)
=-\frac{1}{4\pi}  \varepsilon_{ijk} \partial_i \bar \theta_{jk}(s)
=-\frac{1}{2\pi} \sum_i \sum_{j<k}  \varepsilon_{ijk} 
\{\bar \theta_{jk}(s+\hat i)-\bar \theta_{jk}(s)\}$. 
Hence, the total sum of six $|\bar \theta_{ij}(s)| (i<j) $
is to exceed $2 \pi$ to realize $k_0(s)=\pm 1$.
%
Since large $|\bar \theta_{ij(s)}|$ accompanying large $S^{\rm Abel}$ is 
not preferable,  the magnetic field $|\bar \theta_{ij}|$ around the 
monopole is estimated as 
$| \bar \theta_{ij} | \simeq  2 \pi / 6 = \pi / 3$ on the average, 
using the spherical symmetry of the magnetic field 
in the vicinity of the monopole. 
Accordingly, we estimate as 
$S^{\rm Abel}_{ij} = 1-{\rm  cos}(| \bar \theta_{ij} |)
\simeq 1- \cos \frac{\pi}{3} = \frac12$ 
around the monopole on the average. 
The above argument can be easily generalized to the case with 
arbitrary monopole-current direction. 

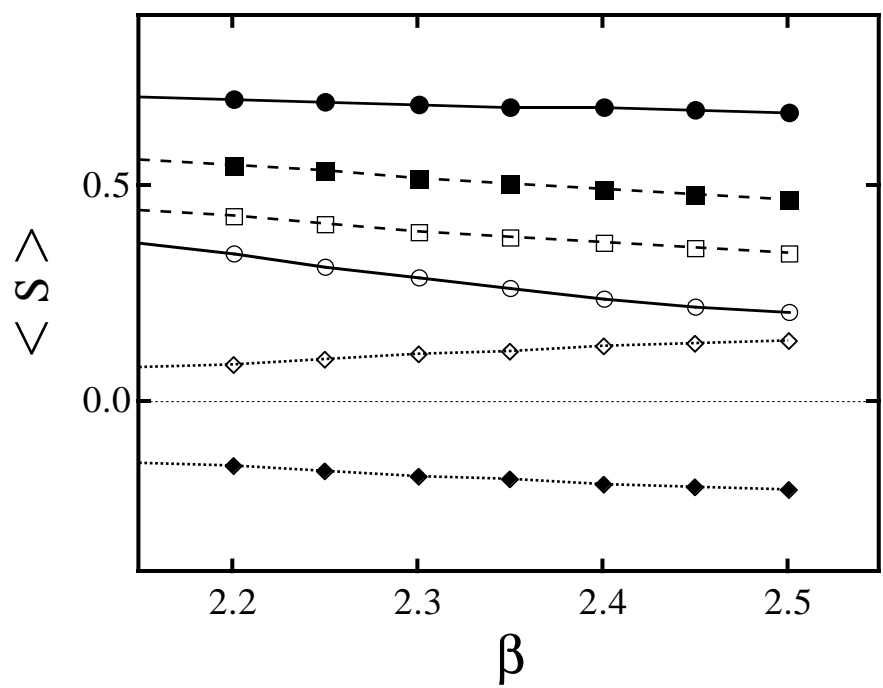
\begin{figure}[t]
\caption{
The action density as the function of $\beta$
in the MA gauge in the SU(2) lattice QCD.
The closed symbols denote the action densities $\langle  S \rangle$
around the monopole current, while the open symbols denote those on 
the whole lattice.
The square, circle and rhombus denote  $\langle S^{\rm SU(2)} \rangle$,
$\langle  S^{\rm Abel} \rangle$ and $\langle S^{\rm off} \rangle$,
respectively.
The monopole  accompanies a large U(1)$_3$ plaquette action,
however, such a large U(1)$_3$ action is 
strongly canceled by the off-diagonal part.
}
\label{gfig10c}
\vspace{0cm}
\end{figure}

Then, existence of monopoles brings a peak 
around $S^{\rm Abel} = \frac12 $ in the distribution $P(S^{\rm Abel})$. 
In fact, the abelian action density $S^{\rm Abel}$ has two ingredients; 
one is nontrivial large fluctuation about $S^{\rm Abel}=1/2$ 
originated from the monopole, and the other is 
remaining small fluctuations, which is expected to vanish 
as $S^{\rm Abel} \rightarrow 0$ as $a \rightarrow 0$.
As shown in Fig.\ref{gfig10c}, the peak originated from
 the monopole is almost $\beta$ independent,
while the other fluctuation becomes small 
for large $\beta$.
At a glance from this result, 
the monopole seems hard to exist at the small mesh $a$, 
since the monopole needs a large abelian action $S^{\rm Abel}$. 
Nevertheless, the monopole can exist in QCD 
even in the large $\beta$ region 
owing to the contribution of the off-diagonal gluon. 
As shown in Fig.\ref{gfig10}(b), 
the off-diagonal part $S^{\rm off}$ of the action density 
around the monopole tends to take a large negative value, and 
strongly cancels with the large abelian action $S^{\rm Abel}$ 
to keep the total SU(2) action $S^{\rm QCD}$ finite.

Here, we consider the angle variable $\tilde \chi_\mu(x)$ of 
the off-diagonal gluons $A_\mu^\pm(x)$ around the monopole. 
In the MA gauge, the amplitude of $A_\mu^\pm(x)$ is strongly reduced, 
and $\tilde \chi_\mu(x)$ can be approximated as a random variable 
on the whole, because $\tilde \chi_\mu(x)$ is free from 
the MA gauge condition entirely and is less constrained from 
the QCD action due to the small $|A_\mu^\pm(x)|$. 
However, around the monopole, 
the off-diagonal gluon $A_\mu^\pm(x)$ inevitably has 
a large amplitude even in the MA gauge to cancel the large 
abelian action density. 
This requirement on the reduction of the total action density 
severely constrains the randomness of the angle variable 
$\tilde \chi_\mu(x)$ of the off-diagonal gluon $A_\mu^\pm(x)$ 
around the monopole. 
As the result, the randomness of $\tilde \chi_\mu(x)$ is weaken, and 
continuity of $\tilde \chi_\mu(x)$ or $A_\mu^\pm(x)$ becomes clear 
in the vicinity of the monopole even in the MA gauge. 
This continuity of $A_\mu^\pm(x)$ around the monopole 
ensures the topological stability 
of the monopole itself as $\Pi_2({\rm SU}(2)/{\rm U}(1))={\bf 
Z}_{\infty}$.

To summarize, existence of the monopole  inevitably accompanies a large 
abelian plaquette action $S^{\rm Abel}$ around it, however, the 
off-diagonal part $S^{\rm off}$ takes a large negative value around the 
monopole and strongly cancels with $S^{\rm Abel}$ to keep $S^{\rm QCD}$ 
not so large. 
Due to this strong cancellation 
between $S^{\rm Abel}$ and $S^{\rm off}$, 
monopoles can appear in the abelian sector in QCD 
without large cost of the QCD action $S^{\rm QCD}$, which 
controls the generating probability of the gluon configuration. 
The extension of the off-diagonal rich region around the monopole
can be interpreted as the effective size or 
the structure of the monopole, because the abelian gauge theory is 
largely modified inside the QCD-monopole like the 
't~Hooft-Polyakov monopole.

Finally, in this section, let us consider the correlation between 
monopoles and instantons \cite{fukushima} in terms of the 
gluon-field fluctuation. 
The instanton is a nontrivial classical solution of 
the Euclidean Yang-Mills theory, 
corresponding to the homotopy group 
$\Pi_3({\rm SU}(N_c)) = {\bf Z}_\infty $\cite{diakonov,shuryak}. 
For the instanton, the SU(2) structure of the gluon field 
is necessary at least. 
In spite of the difference on the topological origin, 
recent studies indicate the strong correlation 
between monopoles and instantons in the QCD 
vacuum in the MA gauge\cite{suganuma3,brower}.
What is the origin of the relation between two different 
topological objects, monopoles and instantons ? 
In the MA gauge, off-diagonal components are forced to be small, 
and the gluon field configuration seems abelian on the whole. 
However, even in the MA gauge, off-diagonal gluons largely remain 
around the QCD-monopole. 
The concentration of off-diagonal gluons around monopoles leads to 
the local correlation between monopoles and instantons: 
instantons appear around the monopole world-line in the MA gauge, 
because instantons need full SU(2) gluon components for existence.

\section{Gluonic Higgs Field in QCD and Monopoles}

QCD in the maximally abelian (MA) gauge has several similarities with 
the nonabelian Higgs (NAH) theory 
in terms of the gauge-symmetry reduction and the appearance 
of monopoles, although the symmetry reduction is realized 
by the gauge fixing instead of Higgs condensation. 
In this section, we try to formulate QCD in the MA gauge 
in the similar manner to the NAH theory. 
To this end, we introduce the concept of the `gluonic Higgs field' 
$\vec \phi_D[A_\mu(x)]$ with $\phi_{Dj} [A_\mu(x)] \in su(N_c) $, 
a gluonic composite scalar defined from the ${\rm SU}(N_c)$ 
covariant derivative $\hat D_\mu$ in subsection 5.1. 
By way of $\vec \phi_D[A_\mu(x)]$, 
we formulate the abelian projection in QCD without explicit use of 
the notion of the gauge fixing. In this formalism, 
the abelian projection resembles the extraction of the photon field 
in the NAH theory. 
In subsection 5.2, we study the connection of 
the gluonic Higgs field with the MA gauge, and 
examine the correspondence between 
the monopole in the MA gauge and the hedgehog configuration of 
$\vec \phi_D[A_\mu(x)]$ in the lattice QCD. 
%

\subsection{Gluonic Higgs Field for the Relevant Abelian 
Submanifold in QCD}

The abelian dominance in the MA gauge observed in the lattice QCD 
indicates the existence of the infrared-relevant abelian gauge 
submanifold embedded in QCD, and 
we call it the `relevant abelian submanifold' in QCD. 
The ordinary abelian projection in the MA gauge can be interpreted as 
an concrete procedure to extract this abelian manifold 
in QCD. Here, we extend the concept of the `abelian projection' as 
the extraction of this relevant abelian submanifold in QCD. 
In the NAH theory, the extraction of the photon field 
corresponds to the abelian projection, and 
the Higgs field $\phi(x)$ indicates the `abelian direction' 
in the nonabelian gauge manifold. 
Based on the similarity with the NAH theory, 
we introduce the `gluonic Higgs field' $\phi [A_{\mu}(x)]$ 
to extract the relevant abelian submanifold in QCD, 
referring the abelian projection in the MA gauge. 
As was shown in section 2, 
if $ \phi[A_{\mu}(x)]$ obeys the adjoint transformation, 
the monopole in the abelian gauge can be understood as 
the topological defect on $\phi [A_{\mu}(x)]$, and 
the abelian-projection scheme in QCD becomes analogous to the 
NAH theory, by regarding $\phi [A_{\mu}(x)]$ as the Higgs field. 

The variable $\Phi[A_{\mu}(x)]$ 
in Eq.(\ref{eq:operator}) appearing in the MA gauge 
seems a candidate of the gluonic Higgs field $\phi[A_\mu(x)]$, 
however, it does not obey the adjoint transformation, 
and correspondence between $ \Phi[A_{\mu}(x)]$ 
and the location of the monopole is unclear. 
In general, when a variable $O(x)$ is diagonalized, all the functions 
$f(O(x))$ are also diagonalized. 
Hence, there appears the ambiguity to choose $\phi[A_{\mu}(x)]$.
Then, we require the following properties for the gluonic Higgs field 
$\phi[A_{\mu}(x)]$ in QCD. 
\begin{enumerate}
\item
To extract $(N_c-1)$ abelian gauge fields $\vec A_\mu(x)$ corresponding 
to U(1)$^{N_c-1} \subset$ SU$(N_c)$, 
the gluonic Higgs field $\vec \phi[A_{\mu}(x)]$ consists of 
$(N_c-1)$ components $\phi_j[A_\mu(x)]$ ($j=1,.., N_c-1$).
Each $\phi_j$ is defined to be an hermite gluonic 
composite scalar as $\phi_j[A_{\mu}(x)]=\phi_j^a T^a \in su(N_c)$ 
with $\phi_j^a \in {\bf R}$. 
\item
Similar to the Higgs field in the NAH theory, 
$\vec \phi [A_{\mu}(x)]$ obeys the adjoint gauge transformation as 
$\vec \phi \rightarrow \vec \phi^\Omega = \Omega
\vec \phi \Omega^\dagger$ by the SU($N_{c}$) gauge 
transformation $\Omega (x) \in$ SU($N_{c}$). 
\item
Corresponding to the direct product of U(1)$^{N_c-1}$, 
the $(N_c-1)$ components $\phi_j[A_\mu(x)] \in su(N_c)$ are to be 
commutable each other as $[\phi_i(x),\phi_j(x)]=0$, 
and are normalized as 
${\rm tr} \{\phi_i(x)\phi_j(x)\} = \frac12 \delta_{ij}$. 
This means that $\vec \phi[A_\mu(x)]$ forms the Cartan subalgebra 
at each local point $x_\mu$, 
and $\vec \phi[A_\mu(x)]$ is required to be written as 
$\vec \phi[A_\mu(x)]= \Omega_C \vec H \Omega_C^{\dagger}$ 
with a suitable $\Omega_C[A_\mu(x)] \in {\rm SU}(N_c)$.
\item
When the gluonic Higgs field $\vec \phi [A_{\mu}(x)]$ 
is diagonalized by the SU$(N_c)$ gauge transformation, 
the diagonal gluon component $\vec A_\mu(x)$ 
is required to provide the relevant abelian submanifold, 
which corresponds to the abelian projected QCD in the MA gauge. 
\end{enumerate}

Inspired from the argument of the GMA gauge, 
the gluonic Higgs field $\vec \phi_D [A_{\mu}(x)]$ 
satisfying the above requirements is found as 
\begin{eqnarray} 
\vec \phi_D [A_{\mu}(x)]  & \hspace{0.5cm} \mbox{s.t.}
\hspace{0.5cm}   & {\rm Min}_{\vec \phi(x) \in C}
R_{\vec \phi}[A_{\mu}(x)] \\ \nonumber 
&  & = {\rm Min}_{\vec \phi(x)\in C} 
\int d^4x {\rm tr}[\hat D_{\mu}, \vec \phi(x)]^\dagger 
[\hat D_{\mu}, \vec \phi(x)], 
\label{eq:qcdhiggs}
\end{eqnarray} 
where ${\rm Min}_{\vec \phi(x) \in C} F[\vec \phi(x)]$ means 
the minimization of $F[\vec \phi(x)]$ by taking a suitable 
$\vec \phi(x) \in C$. 
Here, $C$ denotes the set of the Cartan-frame field $\vec \phi(x)$ 
introduced in section 3.3, and 
$\vec \phi(x)$ can be described as 
$\vec \phi(x)= \Omega_C \vec H \Omega_C^{\dagger}$ with 
$\Omega_C(x) \in {\rm SU}(N_c)$.

The gluonic Higgs field $\vec \phi_D [A_{\mu}] \in su(N_{c})$ 
in Eq.(\ref{eq:qcdhiggs}) is determined directly from the gluon 
configuration $A_{\mu}(x)$, without the notion of the gauge fixing. 
From Eq.(\ref{eq:qcdhiggs}), 
the gluonic Higgs field $\vec \phi_D[A_\mu(x)]$ is interpreted as the 
local `color-direction' averaged over the four SU($N_c$) covariant 
derivatives $\hat D_\mu$, and is a complicated function of the gluon 
field $A_\mu(x)$. 
The local form of the definition of $\vec \phi_D[A_\mu(x)]$ is obtained 
from the extremum condition of $R_{\vec \phi}[A_\mu(x)]$ on 
$\vec \phi(x)$ as
\begin{eqnarray}
[\vec \phi_D (x), 
[\hat D_\mu^\dagger, [\hat D_\mu, \vec \phi_D (x)] ] ]= 0. 
\end{eqnarray}
Hence, $\vec \phi_D [A_\mu(x)]$ is 
derived from $\hat D_\mu$ as the solution of the eigenvalue equation, 
\begin{eqnarray}
[\hat D_\mu^\dagger,[\hat D_\mu, \phi_{Dj} (x)] ] = \lambda_j(x)
\phi_{Dj} (x), 
\end{eqnarray}
where the eigenvalue $\lambda_j(x)$ is non-negative 
due to $\hat D_\mu^{\dagger}=-\hat D_\mu$ and satisfies 
$
R_{\vec \phi_D} [A_\mu(x)] = \frac12 \int d^4x \sum_{j=1}^{N_c-1} 
\lambda_j(x). 
$

As a relevant property, $ \vec \phi_D (x)$ obeys the adjoint 
transformation 
\begin{eqnarray} 
\vec \phi_D [A_{\mu}(x)] \rightarrow 
(\vec \phi_D [A_{\mu}(x)])^V = V(x) 
\vec \phi_D [A_{\mu}(x)] V^\dagger(x)
\end{eqnarray}
by the SU($N_{c}$) gauge transformation with $V$ $\in$ SU($N_{c}$). 
In fact, $R_{\vec \phi_D}[A_{\mu}]$ is transformed as 
\begin{eqnarray} 
R_{\vec \phi_D}[A_{\mu}]  \rightarrow  (R_{\vec \phi_D}[A_{\mu}])^V  
& = & \int d^{4} x \; {\rm tr}
\left( [V  \hat D_{\mu} V^{\dagger}, \vec \phi_D^V]^{\dagger} 
[V\hat D_{\mu} V^{\dagger}, \vec \phi_D^V] \right) \\
& = &  \int d^{4} x \; {\rm tr} 
 \left( [\hat D_{\mu}, V^{\dagger}\vec \phi_D^V V]^{\dagger}
[\hat D_{\mu}, V^{\dagger}\vec \phi_D^V V ]\right),  \nonumber
\end{eqnarray} 
and $(R_{\phi})^V$ is minimized for $\vec \phi^{V}_D = 
V(x) \vec \phi_D(x) V^\dagger (x)$, whose 
uniqueness can be proved by considering 
the infinite product of the infinitesimal gauge transformation. 
In particular, $\vec \phi_D (x)$ is U(1)$^{N_c-1}_{\phi}$ gauge 
invariant, because of $\omega_\phi(x) \vec \phi_D (x) 
\omega_{\phi}^\dagger(x) = \vec \phi_D (x)$ with 
$\omega_{\phi}(x) \equiv e^{i \vec \phi_{D}(x) \cdot \vec \chi(x)}$ 
$\in$ U(1)$_{\phi}^{N_c-1}.$

The gluonic Higgs field $\vec \phi_D [A_{\mu}(x)]$ is introduced to 
indicate the color-direction to be projected and to extract the 
abelian gauge manifold. In this respect, $\vec \phi_D [A_{\mu}(x)]$ 
plays the similar role to the Higgs field in the NAH theory, although
there are 
two following differences.
\begin{enumerate}
\item 
$\vec \phi_D[A_\mu(x)]$ is a composite field of 
the gluon $A_\mu(x)$, and is not an elementary degrees of freedom. 
\item 
$\vec \phi_D [A_\mu(x)]$ only has the color-direction degrees of 
freedom and does not have the amplitude degrees of freedom.  
\end{enumerate}

Now, we consider the projection of the operator $\hat O(x)$ 
to $\vec \phi_D$-direction. 
In the continuum QCD, interesting operators consist of 
the derivative operator and the Lie algebra, which are described by 
the infinitesimal transformation of the corresponding group elements, 
and therefore they are to be expressed with the commutation relation. 
Using the commutation relation with the gluonic Higgs field 
$\vec \phi_D [A_\mu(x)]$, 
we define $\vec \phi_D$-projection of the local infinitesimal 
operator $\hat O(x)$ as 
\begin{eqnarray}
\hat O^{\phi} (x) \equiv \hat O(x) - 
[\vec \phi_D(x), [\vec \phi_D(x), \hat O(x)]], 
\end{eqnarray}
which is a generalized version of Eq.({\ref{eq:APdif}}). 
When $\hat O(x)$ does not include the derivative operator 
$\hat \partial_\mu$, this definition is trivial as 
$\hat O^{\phi}(x)=2{\rm tr}\{\hat O(x) \vec \phi_D(x)\} 
\cdot \vec \phi_D(x)$ for $\hat O(x)=\hat O^a(x)T^a \in su(N_c)$.

The construction of the relevant abelian gauge submanifold in QCD 
is performed with $\vec \phi_D$-projection of 
the SU($N_c$) covariant derivative operator 
$\hat D_\mu \equiv \hat \partial_\mu+ieA_\mu$, 
\begin{eqnarray}
\hat D_\mu^{\phi} \equiv \hat D_\mu - 
[\vec \phi_D, [\vec \phi_D, \hat D_\mu]] 
=\hat \partial_\mu+ieA_\mu^\phi+[\vec \phi_D, \partial_\mu \vec \phi_D] 
\end{eqnarray}
with $A_\mu^\phi(x) 
=2{\rm tr} \{ A_\mu(x) \vec \phi_D(x)\} \cdot \vec \phi_D(x)$. 
Here, the nontrivial term $[\vec \phi_D, \partial_\mu \vec \phi_D]$ 
appears in $\hat D_\mu^\phi$ 
owing to the $x$-dependence of the Cartan-frame field $\vec \phi_D(x)$. 
It is to be noted that 
$[\vec \phi_D, \partial_\mu \vec \phi_D]$ does not include 
$\vec \phi_D$-component as 
\begin{eqnarray} 
{\rm tr} \left( \phi_{Di}(x) 
[\vec \phi_D(x), \partial_\mu \vec \phi_D(x)] \right)
= {\rm tr} \left( \partial_\mu \vec \phi_D(x) [ \phi_{Di}(x), 
\vec \phi_D(x)] \right)=0.
\end{eqnarray}

In this formalism, the abelian projection 
is defined by the replacement of $\hat D_\mu$ by $\hat D_\mu^\phi$. 
Accordingly, the abelian-projected gluon $\tilde A_\mu^\phi(x)$ 
is defined as the difference between $\hat D_\mu^\phi$ and 
$\hat \partial_\mu$, 
\begin{eqnarray}
\tilde A_\mu^\phi(x) \equiv \frac{1}{ie} 
(\hat D_\mu^{\phi}-\hat \partial_\mu)=A_\mu^\phi(x) 
+\frac{1}{ie}[\vec \phi_D(x), \partial_\mu \vec \phi_D(x)] \in su(N_c), 
\end{eqnarray}
and the abelian projection is expressed by 
the mapping of $A_\mu(x) \rightarrow \tilde A_\mu^\phi(x)$. 
It is remarkable that $\tilde A_\mu^\phi(x)$ includes both the 
$\vec \phi_D$-component $A_\mu^{\phi}(x)= 
2{\rm tr}(A_\mu(x)\vec \phi_D(x)) \cdot \vec \phi_D(x)$ 
and the non-$\phi_D$-component 
$\frac{1}{ie}[\vec \phi_D(x), \partial_\mu \vec \phi_D(x)]$. 
The abelian gauge field is defined by the 
$\vec \phi_D$-component of $\tilde A_\mu^\phi(x)$, 
\begin{eqnarray} 
\vec A_\mu^\phi(x) 
\equiv 2 {\rm tr} (\tilde A_{\mu}^{\phi}(x) \vec \phi_D(x)) 
= 2 {\rm tr}(A_\mu^{\phi}(x) \vec \phi_D(x)). 
\end{eqnarray}
Here, $\vec A_\mu^\phi $ is 
the image of $\tilde A_\mu^\phi(x)$ projected into the 
U(1)$_\phi^{N_c-1}$ gauge manifold, and corresponds to the photon 
field in the NAH theory.

As for the gauge symmetry, both 
$\tilde A_\mu^\phi(x)$ and $\vec A_\mu^\phi(x)$ 
surely behave as the abelian gauge field 
under the U$(1)_\phi^{N_c-1}$ abelian gauge-transformation by 
$\omega_\phi(x)=e^{i \vec \phi_D(x) \cdot \vec \chi(x)} 
\in {\rm U}(1)_\phi^{N_c-1}$. 
In fact, $\vec A_\mu^\phi(x)$ is gauge-transformed as 
\begin{eqnarray}
\vec A_\mu^\phi(x) \rightarrow (\vec A_\mu^{\phi_D}(x))^\omega 
&= & 2{\rm tr}(A_\mu^\omega(x) \vec \phi_D^\omega(x))  
 =  2{\rm tr}(\omega \{A_\mu(x)+ \frac{1} {ie} \partial_\mu \}
\omega^\dagger(x) \cdot \vec \phi_D(x))  \nonumber \\ 
& = & 2{\rm tr}(A_\mu(x) \vec \phi_D(x)) + \frac2e {\rm tr}
\{ {\partial_{\mu} (\chi_i(x) \phi_{Di}(x))} \cdot \vec \phi_D (x) \} 
\nonumber \\
& = & \vec A_{\mu}^\phi (x) + {1 \over e} \partial_\mu \vec \chi(x), 
\end{eqnarray}
where we have used 
\begin{eqnarray}
{\rm tr}\{ \partial_\mu   \phi_{Di}(x)  \phi_{Dj}(x) \} & = & 
{\rm tr} \{ \partial_\mu (\Omega _C H_i \Omega_C^\dagger) \cdot 
(\Omega _C H_j \Omega_C^\dagger)\} \nonumber \\  
& = & {\rm tr} ( \Omega_C^\dagger \partial_\mu \Omega _C H_i H_j + 
\partial_\mu \Omega_C^\dagger \Omega_C H_j H_i) = 0.
\end{eqnarray}
Then, $\tilde A_\mu^\phi(x)$ is gauge-transformed as 
\begin{eqnarray}
\tilde A^\phi_\mu(x) \rightarrow (\tilde A^\phi_\mu(x))^\omega
= \tilde A_\mu^{\phi}(x) + 
\frac{1}{e} \partial_\mu \vec \chi_\mu(x) \cdot
\vec \phi_D(x). 
\end{eqnarray}

Next, we study the abelian field strength and the monopole current. 
The abelian field-strength matrix is defined as 
\begin{eqnarray} 
\tilde F_{\mu\nu}^\phi(x) & \equiv & 
\frac{1}{ie} \left( [\hat D_\mu^\phi, \hat D_\nu^\phi]
-[\hat \partial_\mu, \hat \partial_\nu] \right) \nonumber \\
& = & \partial_\mu \tilde A^\phi_\nu(x) - \partial_\nu \tilde 
A^\phi_\mu(x) + ie[\tilde A_\mu^\phi(x), \tilde A_\nu^\phi(x)],
\end{eqnarray}
which generally includes the non-$\vec \phi_D$-component as well as 
$\tilde A_\mu^\phi(x)$. 
The $\vec \phi_D$-component of $\tilde F_{\mu\nu}^\phi(x)$ is 
the image of $\tilde F_{\mu\nu}^\phi(x)$ projected into the 
U(1)$_\phi^{N_c-1}$ gauge manifold, and is 
observed as the `real abelian field-strength' in the 
abelian-projected gauge theory. 
The explicit form of $\vec F_{\mu\nu}^\phi(x)$ is derived as 
\begin{eqnarray}
\vec F_{\mu\nu}^\phi(x)  & \equiv  & 2 {\rm tr}
\left( \tilde F_{\mu\nu}^\phi(x) \vec \phi_D(x) \right)  \nonumber \\
& = & \partial_\mu \vec A_\nu^\phi(x) - 
      \partial_\nu \vec A_\mu^\phi(x) +
      \frac{4}{ie} {\rm tr} (\vec \phi_D(x) [\partial_\mu \phi_{Di}(x), 
      \partial_\nu \phi_{Di}(x)]) \\
& = & \partial_\mu \vec A_\nu^\phi(x) - 
      \partial_\nu \vec A_\mu^\phi(x) + \frac{2}{e} f_{abc}
      \vec \phi_D^a \partial_\mu \phi_{Di}^b \partial_\nu \phi_{Di}^c, 
\end{eqnarray}
where the last term breaks the abelian Bianchi identity and 
provides the monopole current. 
The magnetic monopole current is derived as 
\begin{eqnarray} 
\vec k_\mu^\phi(x) \equiv \partial^{\alpha*} \vec F_{\alpha\mu}^\phi(x) 
= -\frac{1}{e}  \varepsilon_{\mu\alpha\beta\gamma}  f_{abc}
\partial^\alpha \vec \phi^a_{D}(x) \partial^\beta \phi_{Di}^b(x) 
\partial^\gamma \phi_{Di}^c(x), 
\end{eqnarray}
which is the topological current induced by $\vec \phi_D[A_\mu(x)]$. 
Hence, the monopole appears from the center of the hedgehog 
configuration of $\vec \phi_D[A_\mu(x)]$ as shown in Fig.\ref{Higgs} 
in the SU(2) case.


The gluonic Higgs field $\vec \phi_D [A_\mu(x)]$ in QCD 
plays the similar role to the Higgs field in the NAH theory 
on the extraction of the abelian gauge manifold and 
the appearance of the monopole current. 
In principle, the abelian projection can be performed in QCD 
in the gauge-covariant manner 
using the gluonic Higgs field $\vec \phi_D [A_\mu(x)]$ 
without the notion of the gauge fixing.

\subsection{Relation among the Gluonic Higgs Field, 
the MA Gauge Function and Monopoles} 

In this subsection, we consider the correspondence 
among the gluonic Higgs field $\vec \phi_D [A_{\mu}(x)]$ and 
the MA gauge function $\Omega_{\rm MA}[A_{\mu}(x)] \in G/H$, 
and QCD-monopoles. 
As the relation between $\Omega_{\rm MA}$ and $\vec \phi_D$, 
the minimization condition for $R_{\vec H}[A_{\mu}^{\Omega}(x)]$ 
by the gauge degrees of freedom can be equivalently 
rewritten into the minimization of 
$ R_{\vec \phi}[A_{\mu}(x)]$ in terms of $\vec \phi (x) \in C$. 
%
%
Then, the gluonic Higgs field $\vec \phi_D [A_{\mu}(x)] \in C$ 
directly corresponds to the MA gauge function 
$\Omega _{\rm MA}(x)\in G/H$ as 
\begin{eqnarray} 
\vec \phi_D(x) = \Omega_{\rm MA} ^\dagger (x) \vec H \Omega_{\rm MA} (x).
\label{eq:pdiag}
\end{eqnarray}
Then, if the MA gauge is uniquely determined beside 
U(1)$^{N_c-1}_{\rm local} \times$ Weyl$_{N_c}^{\rm global}$, 
$\vec \phi_D [A_\mu(x)]$ is also uniquely determined beside 
the global Weyl symmetry, because of the U$(1)^{N_c-1}$ gauge 
invariance of $\vec \phi_D[A_\mu(x)]$. 

Also from Eq.(\ref{eq:pdiag}), we can derive  
the adjoint gauge-transformation property of the gluonic 
Higgs field $\vec \phi_D [A_\mu(x)]$ again. 
In the arbitrary SU($N_c$) gauge transformation 
with $V(x)\in {\rm SU}(N_c)$, the MA gauge function 
$\Omega _{\rm MA} \in G/H$ obeys the nonlinear gauge 
transformation as 
\begin{eqnarray}
\Omega _{\rm MA}(x) \rightarrow  \Omega _{\rm MA}^V(x)
=d^V(x) \Omega _{\rm MA}(x)V^{\dagger}(x). 
\end{eqnarray}
Here, $d^V(x)\in H$ appears so as to 
keep $\Omega _{\rm MA}^V(x)$ belonging to the coset space $G/H$, i.e., 
$\Omega _{\rm MA}^V(x) \in G/H$. 
Then, $\vec \phi_D (x)$ is transformed by 
$V(x) \in {\rm SU}(N_c)$ as 
\begin{eqnarray}
\vec \phi_D \rightarrow \vec \phi_D^V
=\Omega _{\rm MA}^{V \dagger} \vec H \Omega _{\rm MA}^{V} 
= V \Omega _{\rm MA}^{\dagger} d^{V \dagger}  \vec H d^V 
\Omega_{\rm MA} V^\dagger = V \vec \phi_D V^\dagger, 
\end{eqnarray}
which is nothing but the SU$(N_{c})$ adjoint gauge 
transformation.

Here, we consider the singularity relating to the monopole 
appearing in the abelian gauge manifold of QCD. 
In a suitable gauge like the Landau gauge, 
the gluon field can be taken as a regular field, however, 
the gluonic Higgs field $\vec \phi_D[A_\mu(x)]$ generally 
includes the singularity like the hedgehog configuration 
as shown in Fig.\ref{Higgs} (b), and therefore 
the relevant abelian manifold described by 
$\vec A_{\mu}^\phi \equiv A_{\mu}^a \vec \phi_D^a$ 
holds the monopole singularity 
at the hedgehog center of $\vec \phi_D(x)$. 
On the other hand, in the MA gauge, the gluonic Higgs field 
$\vec \phi_D [A_\mu(x)]$ is arranged into $\vec H$-direction 
by the SU($N_c$) gauge transformation like the unitary gauge 
in the NAH theory. 
Then, $(\vec \phi_D)^{\Omega_{\rm MA}} =
\Omega_{\rm MA} \vec \phi_D \Omega_{\rm MA} ^\dagger 
= \vec H $ becomes trivially regular as shown in Fig.\ref{Higgs}(a). 
Instead, the gluon field $A_\mu^{\rm MA}(x) 
\equiv \Omega^{\rm MA}(x)(A_\mu+\partial_\mu)\Omega^{{\rm MA}\dagger}(x)$ 
includes the singularity as monopoles in the abelian sector.
Such a movement of the singularity from the Higgs field to the gauge field
is also seen during the unitary gauge fixing around the 
't~Hooft-Polyakov monopole in the NAH theory. 

Now, we examine the local correlation between 
the monopole current  and the gluonic Higgs field 
$\vec \phi_D[A_\mu(x)]$ in the SU(2) lattice simulation. 
In the lattice QCD simulation, the gluon configuration 
generated on the lattice is far from the continuous field, 
because of the random appearance of the gauge degrees of 
freedom on each cite. 
On the other hand, to see the topology of $\vec \phi_D[A_\mu(x)]$, 
the gluon field $A_\mu(x)$ is desired to be continuous, 
because discontinuity of $A_\mu(x)$ as the lattice artifact 
inevitably breaks the continuity of $\vec \phi_D(x)$ 
via Eq.(\ref{eq:qcdhiggs}) and provides `fake singularities' 
of $\vec \phi_D(x)$. 
Then, we remove unphysical discontinuity on the gauge degrees of 
freedom from the lattice-QCD gluon configuration 
by a suitable gauge transformation. 
To this end, we impose the SU($N_c$) 
Landau gauge fixing\cite{mandula}, which is defined by maximizing
\begin{eqnarray}
R_L[U_\mu] \equiv {\rm Re} \, \sum_{s,\mu} {\rm tr}  \, U_\mu(s) = N_c 
{\rm Re} \sum_{s,\mu}  U^0_\mu(s)
\end{eqnarray}
using the SU($N_c$) gauge transformation.
In the SU($N_c$) Landau gauge, all the gluon components 
on the lattice become mostly continuous owing to the 
suppression of their fluctuation around $U_\mu(s) = 1$. 
In the continuum limit, this gauge-fixing condition 
coincides the ordinary SU($N_c$) Landau gauge condition, 
$\partial_\mu A_\mu = 0$. 
Thus, we first prepare the continuous gluon configuration 
by the SU(2) Landau gauge fixing, and then the MA gauge fixing 
is performed by the MA gauge function $\Omega^{\rm MA}(x)$. 
The gluonic Higgs field $ \vec \phi_D (x)$ is also 
obtained using Eq.(\ref{eq:pdiag}).

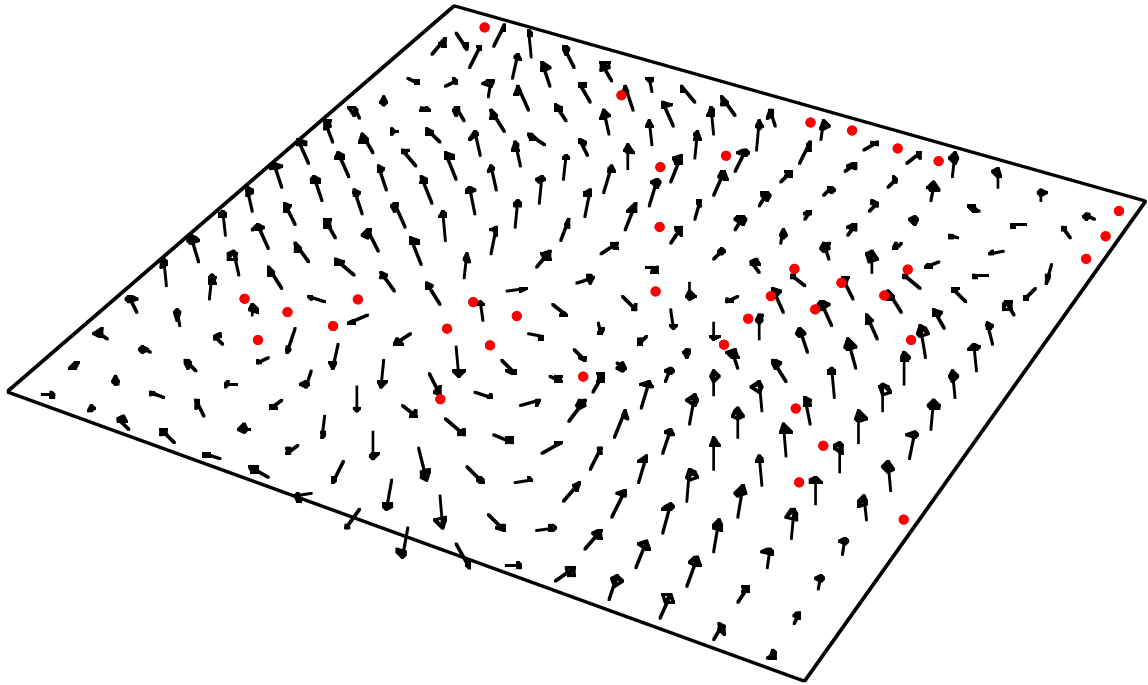
\begin{figure}[t]
\caption{
The local correlation between the gluonic Higgs field 
$\phi_D[A_{\mu}(x)]$ and the monopole in the MA gauge
in the SU(2) lattice QCD with $\beta =2.4$ and $16^4$.
The gluonic Higgs field 
$\phi_D[A_{\mu}(x)] \equiv \phi_D^a \frac{\tau^{a}}{2}$
on a 2-dimensional plane is expressed by the arrow, which denotes its 
color-direction in the SU(2) internal space.
}
\label{phi}
\end{figure}

We show in Fig.\ref{phi} the local correlation between 
the gluonic Higgs field $\phi_D[U_\mu(s)]$ 
and the monopole in the MA gauge 
in the SU(2) lattice QCD with $\beta$=2.4 and $16^4$.
The gluonic Higgs field $\phi_D(s)=\phi_D^a(s)T^a$ 
is expressed by the arrows $(\phi_D^1, \phi_D^2, \phi_D^3)$ 
in the SU(2) internal space. 
The tendency of the local correspondence is found 
between the hedgehog configuration of 
the gluonic Higgs field $\vec \phi_D(x)$ 
and the monopole in the MA gauge.

Finally, we consider the physical meaning of the gluonic Higgs field 
$\vec \phi_D[A_\mu(x)]$ and relevant abelian manifold embedded in QCD. 
Here, the gluonic Higgs field $\vec \phi_D[A_\mu(x)]$ 
can be obtained in the gauge-covariant manner 
using Eq.(\ref{eq:pdiag}) from the gluon field $A_{\mu}(x)$ in QCD, 
and the abelian projection can be performed with $\vec \phi_D(x)$ 
without the notion of the gauge fixing. 
Physically, $\vec \phi_D [A_\mu(x)]$ means the local color-direction 
which is determined so as to minimize the difference between 
the SU$(N_c)$ gauge connection $\hat D_\mu$ and the abelian gauge 
connection $\hat D_\mu^\phi$ along $\vec \phi_D (x)$. 
In terms of the maximal similarity of $\hat D_\mu^\phi$ 
with $\hat D_\mu$, the gluonic Higgs field $\vec \phi_D[A_\mu(x)]$ 
indicates a relevant color-direction peculiar 
to the gluon field $A_\mu(x)$, 
and the projection into the color-direction $\vec \phi_D(x)$ 
provides the extraction of a relevant abelian gauge manifold 
embedded in QCD in the gauge-covariant manner. 

Since the gluonic Higgs field $\vec \phi_D[A_\mu(x)]$ 
obeys the adjoint gauge transformation, it has the direct 
similarity to the Higgs field in the NAH theory. 
Several parallel arguments to the NAH theory 
are applicable for abelian-projected QCD in terms of 
$\vec \phi_D[A_\mu(x)]$ on the extraction of the abelian gauge manifold 
and appearance of the monopole from the hedgehog configuration. 
In terms of the `gluonic Higgs theory' with $\vec \phi_D (x)$, 
the MA gauge fixing directly corresponds to the unitary gauge 
fixing in the NAH theory. 
In particular, abelian dominance in the MA gauge observed 
in the lattice QCD indicates 
that only the $\vec \phi_D$-component gluon remains 
at the long-distance scale like the photon field in the NAH theory, 
and the other gluon component perpendicular to $\vec \phi_D$ becomes 
infrared-irrelevant like the charged massive vector field 
in the NAH theory. 
In other words, the abelian gauge submanifold projected to 
$\vec \phi_D [A_\mu(x)]$ in QCD is considered to hold essence of 
the whole nonabelian gauge manifold in the infrared region. 

\section{Summary and Concluding Remarks}

On the basis of the dual Higgs picture for confinement, 
we have studied the properties of monopoles and gluon fields 
in QCD in the maximally abelian (MA) gauge 
both in the analytical framework and 
in the lattice QCD calculation. 
In the dual Higgs theory, color confinement is realized 
by the one-dimensional squeezing of the color-electric flux in the QCD 
vacuum through the dual Meissner effect caused by monopole condensation. 
The extraction of the abelian gauge theory and 
the appearance of monopoles in QCD can be carried out by taking 
the 't~Hooft abelian gauge, which is defined by diagonalizing 
the `gluonic Higgs field' $\phi[A_{\mu}(x)]$. 

In the abelian gauge, SU($N_c$) gauge theory is reduced into 
U(1)$^{N_c-1}$ gauge theory including the monopole,
which topologically appears corresponding to the nontrivial homotopy 
group, $\Pi_2($SU$(N_c)/$U$(1)^{N_c-1})={\bf Z}^{N_c-1}$. 
In the abelian gauge, the diagonal gluon component 
behaves as the U(1)$^{N_c-1}$ gauge field, while 
the off-diagonal gluon behaves as the charged matter field 
in terms of the residual gauge symmetry. 
For $\phi[A_{\mu}(x)]$ obeying the adjoint gauge transformation, 
the hedgehog configuration of $\phi[A_{\mu}(x)]$ leads to 
the unit-charge magnetic monopole. 
In the abelian gauge, multi-charge monopoles do not appear 
in ${\bf R}^3$ in general cases, because of the over condition.

Using the gauge-connection formalism, the appearance of 
the Dirac string and the monopole has been studied in relation with 
the SU($N_c$) singular gauge transformation. 
The appearance of the Dirac string is originated from 
the multi-valuedness of the gauge function $\Omega(x)$, which leads to 
the divergence of $\Omega^\dagger \partial_\mu \Omega$. 
In the singular SU(2) gauge transformation, 
the multi-valued point of $\Omega(x)$ 
terminates at the hedgehog center of $\phi[A_{\mu}(x)]$, which leads to 
the appearance of the monopole. 
We have shown the relevant role of off-diagonal gluons 
for the appearance of monopoles. 

The maximally abelian (MA) gauge has been well formulated 
in terms of the gauge connection. 
To remove the explicit use of the specific direction as $\vec H$, 
we have formulated the generalized maximally abelian (GMA) gauge 
by introducing the Cartan-frame field $\vec \phi(x)$, which is 
the local Cartan sub-algebra defined at each point. 
The generalization of the abelian projection has been defined 
based on the commutation relation. 
We have investigated the gauge-transformation properties 
of the GMA gauge function $\Omega_{\rm GMA}$, 
and have derived the criterion on the SU($N_c$) gauge 
invariance for the variable $\hat O[A_\mu(x)]^{\Omega_{\rm GMA}}$ 
defined in the GMA gauge. This criterion is also applicable for the 
MA gauge. 

We have studied the gluon-field properties around the monopole 
in QCD in the MA gauge in terms of the field strength and 
the action density both in the analytical consideration 
and in the lattice QCD simulation. 
The monopole provides large field 
fluctuations in the abelian sector: both 
the abelian gauge field and the abelian action density are largely 
fluctuated around monopoles. 
The large fluctuation of off-diagonal gluons has been 
also found around the monopole in the MA gauge, and 
the off-diagonal rich region indicates 
the effective size and the structure of monopoles, 
which is similar to the 't~Hooft-Polyakov monopole.
Since the instanton needs the full SU(2) structure, 
it is expected to appear in the off-diagonal rich region 
around the monopole in the MA gauge, 
which leads to the local correlation between monopoles and instantons. 
We have found the large cancellation between abelian part 
and the off-diagonal part of the action in the MA gauge. 
Owing to this cancellation, the monopole can appear 
in the abelian sector in QCD without large cost of the QCD action, 
although existence of monopoles inevitably enlarges 
the abelian action. 
In other words, the off-diagonal gluon is necessary 
for existence of the monopole in the short-distance scale, 
and the effective monopole size relating to the off-diagonal gluon 
can be regarded as the critical scale of the abelian projected QCD, 
in the similar sense of the correspondence between 
the GUT monopole size and the GUT scale.

Finally, the abelian projection in QCD 
has been formulated without the notion of gauge fixing 
in the similar manner to the extraction of the photon field 
in the nonabelian Higgs theory, 
by introducing the `gluonic Higgs field' 
$\vec \phi_D [A_\mu(x)]$ defined from the gluon field. 
Here, the color-direction 
$\vec \phi_D [A_\mu(x)]$ is determined so as to 
minimize the difference between 
the SU($N_c$) gauge connection and the projected abelian gauge 
connection along $\vec \phi_D [A_\mu(x)]$. 
The gluonic Higgs field $\vec \phi_D[A_\mu(x)]$ obeys the adjoint 
transformation by the SU($N_c$) gauge transformation, 
and the monopole current appears at 
the center of the hedgehog configuration of 
$\vec \phi_D[A_\mu(x)]$ in the SU(2) case.

We would like to thank Professor Yoichiro Nambu for 
his useful comments and discussions. 
One of authors (H.S.) is supported in part by 
Grant for Scientific Research (No.09640359) from 
the Ministry of Education, Science and Culture, Japan.
One of authors (H.I.) is supported by 
Research Fellowships of the Japan Society for the 
Promotion of Science for Young Scientists.
The lattice QCD simulations have been performed on VPP500 at RIKEN and
sx4 at RCNP.

\newpage



\begin{figure}[bt]
\caption{Comparison among QCD,  abelian projected QCD (AP-QCD) and QED
in terms of 
the gauge symmetry and essential degrees of freedom. 
}
\label{APQCD}
\end{figure}

\begin{figure}[bt]
\caption{
Topological structure of the gluonic Higgs field 
$\phi[A_{\mu}(x)]$ 
in the abelian gauge fixing in the SU(2) QCD.
In the abelian gauge, the monopole appears at the singular
point of $\hat \phi(x) \equiv \phi/|\phi|$ with $|\phi| \equiv 
(\phi^a\phi^a)^{1/2}$. 
(a) For the regular (trivial) configuration of $\hat 
\phi[A_{\mu}(x)]$,
 no monopole appears in the abelian gauge.
(b) For the hedgehog configuration of $\hat 
\phi[A_{\mu}(x)]$,
the unit-charge monopole appears in the abelian gauge.
}
\label{Higgs}
\end{figure}

\begin{figure}[bt]
\caption{
Appearance of monopoles in abelian projected QCD(AP-QCD).
After the abelian gauge fixing, monopole with the Dirac string 
appears
from 
$G^{\rm linear}_{\mu\nu} $ 
in Eq.(\ref{eq:field-strength})
and the `anti-Dirac string' appears in the singular part  
$G^{\rm sing}_{\mu\nu}$. 
The off-diagonal contribution $G_{\mu\nu}^{\rm bilinear} = 
ie[A_{\mu},A_{\nu}]$ forms the anti-monopole configuration and 
compensates to the singularity of the other parts.
As the result, the monopole without the Dirac string appears
in the abelian field strength 
${\cal F}_{\mu\nu} $
in AP-QCD. 
}
\label{gfig7}
\end{figure}

\newpage

\begin{thebibliography}{99}

      \bibitem{itzykson}
C.~Itzykson and J.-B.~Zuber, ``Quantum Field Theory'',
(McGraw-Hill, New York, 1985) 1.

      \bibitem{cheng}
For instance, T.~P.~Cheng and L.~F.~Li,
``Gauge Theory of Elementary Particle Physics",
(Clarendon press, Oxford, 1984) 1.

     \bibitem{abrikosov}
A.~A.~Abrikosov,
``Fundamentals of the Theory of Metals",
(North-Holland,  1988) 1.

     \bibitem{diakonov}
D.~I.~Diakonov and V.~Yu.~Petrov, Nucl.~Phys.~{\bf B272}  (1986) 457. 
      
     \bibitem{shuryak}
E.~V.~ Shuryak, Phys. Rep. {\bf 115} (1984) 151.

     \bibitem{hashimoto}
T.~Hashimoto, S.~Hioki, A.~Kawazoe, O.~Miyamura, Y.~Osada, K.~Hirose,
T.~Kanki, M.~Masujima.
Phys. Rev. {\bf D42} (1990) 620. 
 
     \bibitem{diacomo1}
A~.Di~Giacomo, Proc. of ``Quark Confinement and the Hadron Spectrum II'',
(World Sicentific 1997) 64.
         
     \bibitem{thooftM}

G.~'t~Hooft, Nucl. Phys. {\bf B79} (1974) 276.
          
     \bibitem{polyakovM}

A.~M.~Polyakov, JETP Lett. {\bf 20} (1974) 194; Sov. Phys. JETP {\bf 
41} (1976) 988.

     \bibitem{dirac}
P.~M.~A.~Dirac, Proc. of Roy. Soc.{\bf A133} (1931) 60. 

     \bibitem{seiberg}
N.~Seiberg and E.~Witten, Nucl.~Phys.~{\bf B426} (1994) 19;
{\bf B431} (1994) 484.















     \bibitem{nambu}
Y.~Nambu, Phys.~Rev.~{\bf D10} (1974) 4262.


     \bibitem{thoa}
G.~'t~Hooft, ``High Energy Physics'', ed. A.~Zichichi
(Editorice Compositori, Bologna, 1975).

     \bibitem{mandelstam}
S.~Mandelstam, Phys.~Rep.~{\bf C23} (1976) 245.

      \bibitem{rothe}
H.~J.~Rothe, ''Lattice Gauge Theories'', (World Scientific, 1992) 1. \\
M.~ Creutz, ''Quarks, Gluons and Lattices'' (Cambridge, 1983) 1.

     \bibitem{haymaker}
Y.~Peng and R.~W.~Haymaker, Nucl.~Phys.~{\bf B} (Proc. Suppl.) {\bf 34} (1996)
266. \\
R.~W.~Haymaker, V.~Singh, Y.~Peng and J.~Wosiek,
Phys. Rev.{\bf D53}~(1996)~389.

     \bibitem{thooft} 
G.~'t~Hooft, Nucl.~Phys.~{\bf B190} (1981) 455.


     \bibitem{ezawa}
Z.~F.~Ezawa and A.~Iwazaki,
Phys.~Rev.~{\bf D25} (1982) 2681; {\bf D26} (1982) 631.


     \bibitem{maedan}
S.~Maedan and T.~Suzuki, Prog.~Theor.~Phys. {\bf 81} (1989) 229. \\
T.~Suzuki, Prog.~Theor.~Phys. {\bf 80} (1988) 929;
{\bf 81} (1989) 752. \\
S.~Maedan, Y.~Matsubara and T.~Suzuki,
Prog.~Theor.~Phys.~{\bf 84} (1990) 130. 


     \bibitem{suganuma}
H.~Suganuma, S.~Sasaki and H.~Toki,
Nucl.~Phys.~{\bf B435} (1995) 207. 

     \bibitem{sasaki}
S.~Sasaki, H.~Suganuma and H.~Toki,
Prog.~Theor.~Phys.~{\bf 94} (1995) 373.  \\
H.~Suganuma, S.~Umisedo, S.~Sasaki, H.~Toki and O.~Miyamura,
Aust.~J.~Phys. {\bf 50} (1997) 233.  

     \bibitem{umisedo}
S.~Umisedo, H.~Suganuma and H.~Toki,   
Phys. Rev. {\bf D57} (1998) 1605.

     \bibitem{ichie6}
H.~Ichie and H.~Suganuma, 
Proc. of Int. Workshop on ``Future Directions
in Quark Nuclear Physics'', 
Adelaide, Australia, Mar. 1998, (World Scientific), in press;
hep-lat/9807006.

    \bibitem{ichie1}
 H.~Ichie, H.~Suganuma and H.~Toki, 
Phys.~Rev.~{\bf D54} (1996) 3382; {\bf D52} (1995) 2994. \\ 
H.~Monden,  H.~Ichie,  H.~Suganuma and H.~Toki
Phys.~Rev.~{\bf C57}  (1998) 2564.

      \bibitem{ichie2}
H.~Ichie,  A. Tanaka and H. Suganuma,
Nucl.~Phys.~{\bf B} (Proc.~Suppl.) {\bf 63A-C} (1998) 468. \\
H. Ichie, H. Suganuma and A. Tanaka 
Nucl.~Phys.~{\bf A629} (1998) 82c. 

     \bibitem{atanaka}
A.~Tanaka and H.~Suganuma, Proc. of Int. Symp. 
on ``Innovative
Computational Methods in Nuclear Many-Body Problems'', Osaka, Nov. 1997, 
(World Scientific) in press; hep-lat/9712027. 


      \bibitem{kondo}
K.-I.~Kondo, Phys.~Rev.~{\bf D57} (1998) 7467; hep-th/9805153.


      \bibitem{diacomo}
A.~Di~Giacomo,  Nucl.~Phys.~{\bf B} (Proc. Suppl.) {\bf 47} (1996) 136
and references therein.    

     \bibitem{poly}
M.~I.~Polikarpov, Nucl.~Phys.~{\bf B} (Proc. Suppl.) {\bf 53} (1997) 134
and references therein.



     \bibitem{ichiead}
H.~Ichie and H.~Suganuma, preprint, hep-lat/9807025.

     \bibitem{suganuma1}
H. Suganuma, H. Ichie, A. Tanaka and  K. Amemiya,
Prog. Theor. Phys. Suppl. {\bf 131} (1998) in press;
hep-lat/9804027, and references therein.

    \bibitem{yotsuyanagi}
T.~Suzuki and I.~Yotsuyanagi, 
Phys.~Rev.~{\bf D42} (1990) 4257. 

     \bibitem{hioki}
Nucl.~Phys.~{\bf B} (Proc.~Suppl.) {\bf 20} (1991) 236; 
S.~Hioki, S.~Kitahara, S.~Kiura, Y.~Matsubara, 
O.~Miyamura, S.~Ohno and T.~Suzuki, 
Phys.~Lett.~{\bf B272} (1991) 326. \\
S.~Hioki, S.~Kitahara, S.~Ohno, T.~Suzuki, Y.~Matsubara 
and O.~Miyamura, Phys.~Lett.~{\bf B285} (1992) 343.

      \bibitem{bali}
G.~S.~Bali, V.~Bornyakov,  M.~Muller-Preussker and K.~Schilling,
Phys. Rev. {\bf D54}~(1996)~2863.

     \bibitem{miyamura}
O.~Miyamura, Phys.~Lett.~{\bf B353} (1995) 91;
Nucl.~Phys.~{\bf B} (Proc. Suppl.) {\bf 42} (1995) 538. \\
O.~Miyamura and S.~Origuchi,
``Confinement '95", (World Scientific, 1995) 65.


      \bibitem{woloshyn}
R.~M.~Woloshyn, Phys.~Rev.~{\bf D51} (1995) 6411.\\
F.~X.~Lee, R.~M.~Woloshyn and H.~D.~Trottier,
Phys.~Rev.~{\bf D53} (1996) 1532.


     \bibitem{kronfeld}
A.~S.~Kronfeld, D.~Schierholz and U.-J.~Wiese, Nucl.~Phys. {\bf B293}
(1987) 461. \\
A.~S.~Kronfeld, M.~L.~Laursen, G.~Schierholz and  U.~J.~Wiese, 
Phys.~Lett.~{\bf 198B} (1987) 516. 

     \bibitem{schierholz}    
F.~Brandstater, U.~J.~Wiese and G. Schierholz,
Phys.~Lett.~{\bf B272} (1991) 319.

      \bibitem{ichiec}
H.~Ichie and H.~Suganuma, Proc. of INSAM Symp. '96, Hiroshima, 
INSAM report, hep-lat/9709109.

     \bibitem{suganuma4}
H.~Suganuma, M.~Fukushima, H.~Ichie and A.~Tanaka,
Nucl.~Phys.~{\bf B} (Proc. Suppl.) {\bf 65 }(1998) 29. 

     \bibitem{suganuma2}
H.~Suganuma, K.~Itakura and H.~Toki, preprint, hep-th/9512141. 

     \bibitem{degrand}
T. DeGrand and D. Toussaint, Phys. Rev. {\bf D22} (1980) 2478.


     \bibitem{poulis}
G.~I.~Poulis,  Phys.~Rev.~{\bf D54} (1996) 6974.

     \bibitem{georgi}
H.~Georgi, ``Lie Algebras in Particle Physics'', (Benjamin/Cummings, 
1982) 1.  

      \bibitem{ichiem}
H.~Ichie and H.~Suganuma, Proc. of Int. Symp. 
on ``Innovative
Computational Methods in Nuclear Many-Body Problems'', Osaka, Nov. 1997, 
(World Scientific),  hep-lat/9802032.

     \bibitem{mandula}
J.~E.~Mandula and M.~Ogilvie, Phys.~Lett.~{\bf B185} (1987) 127.


     \bibitem{amemiya}
K.~Amemiya and H.~Suganuma, Proc. of Int. Symp. 
on ``Innovative
Computational Methods in Nuclear Many-Body Problems'', Osaka, Nov. 1997, 
(World Scientific) in press; hep-lat/9712028. 




     \bibitem{fukushima}
M.~Fukushima,  S.~Sasaki, H.~Suganuma, A.~Tanaka, H.~Toki and D.~Diakonov,
Phys.~Lett. {\bf B399} (1997) 141, \\
M.~Fukushima, A.~Tanaka, S.~Sasaki, H.~Suganuma, H.~Toki and D.~Diakonov,
Nucl.~Phys.~{\bf B} (Proc.~Suppl.) {\bf 53} (1997) 494.

     \bibitem{suganuma3}
H.~Suganuma, S.~Sasaki, H.~Ichie, F.~Araki and O.~Miyamura,
Nucl.~Phys.~{\bf B} (Proc. Suppl.) {\bf 53} (1997) 528.

     \bibitem{brower}
R.~C.~Brower, K.~N.~Orginos and C.~I.~Tan,    
Phys.~Rev.~{\bf D55} (1997) 6313;
Nucl.~Phys.~{\bf B} (Proc. Suppl.) {\bf 53} (1997) 488.




    








\end{thebibliography}
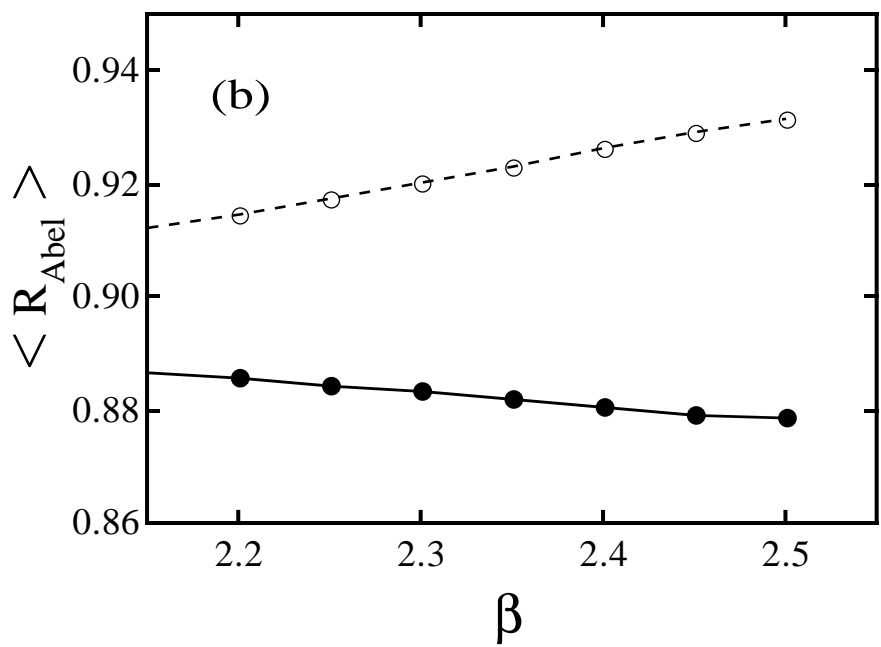
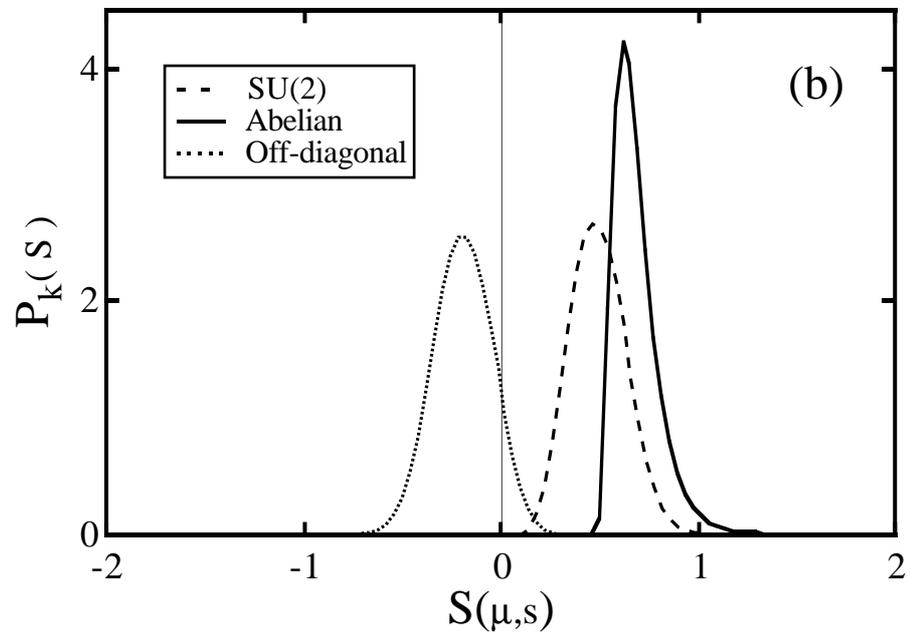
\end{document}